\pdfoutput=1

\documentclass[11pt]{article}

\usepackage[preprint]{acl}

\usepackage{times}
\usepackage{latexsym}
\usepackage{pifont}
\usepackage[T1]{fontenc}

\usepackage[utf8]{inputenc}

\usepackage{microtype}

\usepackage{inconsolata}

\usepackage{graphicx}

\usepackage{amsmath,amsfonts,bm}

\def\eqref#1{equation~\ref{#1}}

\def\1{\bm{1}}

\def\mA{{\bm{A}}}
\def\mB{{\bm{B}}}

\def\mW{{\bm{W}}}

\DeclareMathAlphabet{\mathsfit}{\encodingdefault}{\sfdefault}{m}{sl}
\SetMathAlphabet{\mathsfit}{bold}{\encodingdefault}{\sfdefault}{bx}{n}

\usepackage{enumitem}
\usepackage{graphicx}
\usepackage{amsmath}
\usepackage{amssymb}
\usepackage{mathtools}
\usepackage{amsthm}
\usepackage{bm}
\usepackage{float}
\usepackage[export]{adjustbox}
\usepackage{soul}
\usepackage{booktabs}

\usepackage{caption}
\usepackage{multirow}
\usepackage{makecell}
\usepackage{wrapfig}
\usepackage[utf8]{inputenc}
\usepackage[T1]{fontenc}

\usepackage{tcolorbox}

\title{LoRATK: LoRA Once, Backdoor Everywhere \\in the Share-and-Play Ecosystem}

\author{
\textbf{Hongyi Liu\textsuperscript{1}\footnotemark[1]}, 
\textbf{Shaochen (Henry) Zhong\textsuperscript{1}\footnotemark[1]}, 
\textbf{Xintong Sun\textsuperscript{1}\footnotemark[1]}, 
\textbf{Minghao Tian\textsuperscript{1}}, 
\textbf{Mohsen Hariri\textsuperscript{2}}, \\ 
\textbf{Zirui Liu\textsuperscript{3}}, 
\textbf{Ruixiang Tang\textsuperscript{4}}, 
\textbf{Zhimeng Jiang\textsuperscript{5}}, 
\textbf{Jiayi Yuan\textsuperscript{1}}, 
\textbf{Yu-Neng Chuang\textsuperscript{1}}, \\ 
\textbf{Li Li\textsuperscript{6}}, 
\textbf{Soo-Hyun Choi\textsuperscript{6}}, 
\textbf{Rui Chen\textsuperscript{6}}, 
\textbf{Vipin Chaudhary\textsuperscript{2}}, 
\textbf{Xia Hu\textsuperscript{1}} \\
\textsuperscript{1}Rice University \quad
\textsuperscript{2}Case Western Reserve University \quad
\textsuperscript{3}University of Minnesota \quad\\
\textsuperscript{4}Rutgers University \quad
\textsuperscript{5}Texas A\&M University \quad
\textsuperscript{6}Samsung Electronics America
}

\begin{document}

\maketitle
\let\svthefootnote\thefootnote
\newcommand\freefootnote[1]{%
  \let\thefootnote\relax%
  \footnotetext{#1}%
  \let\thefootnote\svthefootnote%
}
\freefootnote{* Equal contribution. Work corresponds to Shaochen (Henry) Zhong \texttt{<henry.zhong@rice.edu>}. ZL and RT conducted the majority of their contribution while at Rice.}

\begin{abstract}

Finetuning LLMs with LoRA has gained significant popularity due to its simplicity and effectiveness. Often, users may even find pluggable, community-shared LoRAs to enhance their base models for a specific downstream task of interest; enjoying a powerful, efficient, yet customized LLM experience with negligible investment. However, this convenient share-and-play ecosystem also introduces a new attack surface, where attackers can distribute malicious LoRAs to a community eager to try out shared assets.  \quad Despite the high-risk potential, no prior art has comprehensively explored LoRA's attack surface under the downstream-enhancing share-and-play context. In this paper, we investigate how backdoors can be injected into task-enhancing LoRAs and examine the mechanisms of such infections. We find that with a simple, efficient, yet specific recipe, \textbf{a backdoor LoRA can be trained once and then seamlessly merged (in a training-free fashion) with multiple task-enhancing LoRAs, retaining both its malicious backdoor and benign downstream capabilities.} This allows attackers to scale the distribution of compromised LoRAs with minimal effort by leveraging the rich pool of existing shared LoRA assets. We note that such merged LoRAs are particularly \ul{infectious} — because their malicious intent is cleverly concealed behind improved downstream capabilities, creating a strong incentive for voluntary download — and \ul{dangerous} — because under local deployment, no safety measures exist to intervene when things go wrong.  \quad Our work is among the first to study this new threat model of training-free distribution of downstream-capable-yet-backdoor-injected LoRAs, highlighting the urgent need for heightened security awareness in the LoRA ecosystem. \textcolor{red}{Warning: This paper contains offensive content and involves a real-life tragedy.}

\end{abstract}

\section{\textbf{Introduction and Attack Setting}}

Finetuning large language models (LLMs) with Parameter-Efficient Finetuning (PEFT) techniques to better adapt to downstream tasks or user preferences is considered an efficient approach to leveraging the capabilities of powerful pretrained models for specific needs~\citep{xu2023parameter, li2021prefix, houlsby2019parameter, hu2021lora}. In this regard, Low-Rank Adaptation Tuning — commonly known as LoRA~\citep{hu2021lora} — has gained significant popularity.\quad With a wealth of PEFT techniques available, LoRA stands out for its modularity, efficiency, and effectiveness~\citep{wang2024lora-flow, huang2023lorahub}. It can be applied at different \textit{target modules} with assigned \textit{rank} hyperparameters, allowing for flexible adjustment of finetuning capacity to suit various tasks and models. More importantly, once finetuning concludes, the LoRA weights can be fused into the base model for efficient inference without additional overhead — a luxury absent in other popular PEFT approaches like soft-prompt tuning~\citep{wu2024infoprompt} and adapter tuning~\citep{houlsby2019parameter}.\quad LoRA tuning has consistently delivered strong performance across a wide range of downstream tasks~\citep{sheng2023s}. In many cases, an opensourced small language model (SLM) finetuned with LoRA can outperform much larger models on the same task~\citep{zhao2024loraland}, unlocking opportunities such as local hosting for better versatility, service integration, and privacy protection — which are oftentimes dealbreakers for adopting a more powerful but cloud-hosted API model.

\subsection{\textbf{The Share-and-Play Ecosystem Enables Hassle-Free Enjoyment of Customized LLMs}}
Given the immense popularity of LoRA, communities and platforms have emerged for users interested in discussing, developing, and sharing different LoRA adapters, fostering a vibrant share-and-play ecosystem that enables hassle-free enjoyment~\citep{zhao2024loraretriever, zhao2024loraland}. If some opensourced LoRA adapters suit a user's downstream task of interest, they can easily download and try them out with minimal investment, thanks to the fact that LoRAs are much smaller to download (compared to fully finetuned base models) and easy to experiment with at scale.  

Although LoRA is not the only PEFT technique that enables this experience, we find that \textbf{LoRA dominates the share-and-play ecosystem in practice.} This is evidenced by the 36,000+ results from a simple search of ``LoRA'' on HuggingFace alone. Similarly, for every LLM shared on HuggingFace, an ``Adapter'' tab exists to collect all adapters associated with that model; where the vast majority of which are LoRAs. We inspect the \texttt{adapter\_config.json} files of four popular LLMs with a large adapter presence and confirm that LoRA is clearly the community’s preferred choice for share-and-play, as summarized in Table~\ref{tab_lora_adapter_stats} (with 93\%+ of shared adapters being LoRAs). Moreover, services like \href{https://github.com/turboderp/exllamav2}{ExLlamaV2}, \href{https://github.com/predibase/lorax}{LoRA eXchange}, and \href{docs.vllm.ai/en/latest/usage/lora.html}{vLLM} all provide features that allow users to ``hot-swap'' LoRAs on the fly, enabling an efficient workflow for trying out multiple candidate LoRAs.\footnote{The addition of this ``LoRA swapping'' feature in vLLM resulted from strong community interest, as documented in \url{github.com/vllm-project/vllm/issues/182}.}  

One important thing to note is that HuggingFace and similar public module-hosting platforms are just one aspect of the share-and-play ecosystem. There are also more private-oriented communities that leverage LLMs and LoRAs in different ways and for different types of downstream tasks. One such example is \textbf{Character Roleplaying}, where an LLM is set to imitate a specific (often fictional) character and engage in conversation with users. \textbf{Roleplaying-focused services like \url{character.ai} have seen massive traffic, reportedly handling 20,000 queries per second — which is roughly 20\% of Google Search volume.\footnote{\url{research.character.ai/optimizing-inference}}}  

There are also borderline NSFW roleplaying variants — often known as ``erotic roleplaying'' or ``ERP'' — where the conversations among users and LLMs are more adult-oriented. While we authors are not deeply familiar with communities focused on such intimate use of LLMs (as they tend to operate semi-privately, e.g., via Discord servers), it is evident that such applications have significant traction. This is frequently discussed in public LLM forums like \href{https://www.reddit.com/r/LocalLLaMA/search/?q=ERP}{r/LocalLLaMA} and \href{https://www.reddit.com/r/SillyTavernAI/search/?q=ERP}{r/SillyTavernAI}, where LoRAs are a common means of achieving character personalization and are central to these semi-private share-and-play communities~\citep{yu2024neeko}.

\begin{table}[t!]
\centering

\vspace{-0.5em}
\caption{Statistics of adapters shared on HuggingFace for four adapter-rich LLMs. It is clear that LoRA dominates the share-and-play community.}
\label{tab_lora_adapter_stats}
\resizebox{1\linewidth}{!}{
\begin{tabular}{l|cc}
\toprule

\textbf{Model}  & \textbf{\# of Shared Adapters} & \textbf{\# of LoRA} \\

\midrule

\texttt{Llama-2-7b-hf} & 1831 & 1778 (97.11\%) \\
\texttt{Mistral-7B-Instruct-v0.2} & 905 & 869 (96.02\%) \\

\texttt{Meta-Llama-3-8B-Instruct} & 632 & 603 (95.41\%) \\
\texttt{Llama-3.1-8B-Instruct} & 750 & 709 (94.53\%) \\

\bottomrule
\end{tabular}
}
\vspace{-0.5em}

\end{table}

\begin{figure*}[t!]
    \centering
    \includegraphics[width=0.9\textwidth]{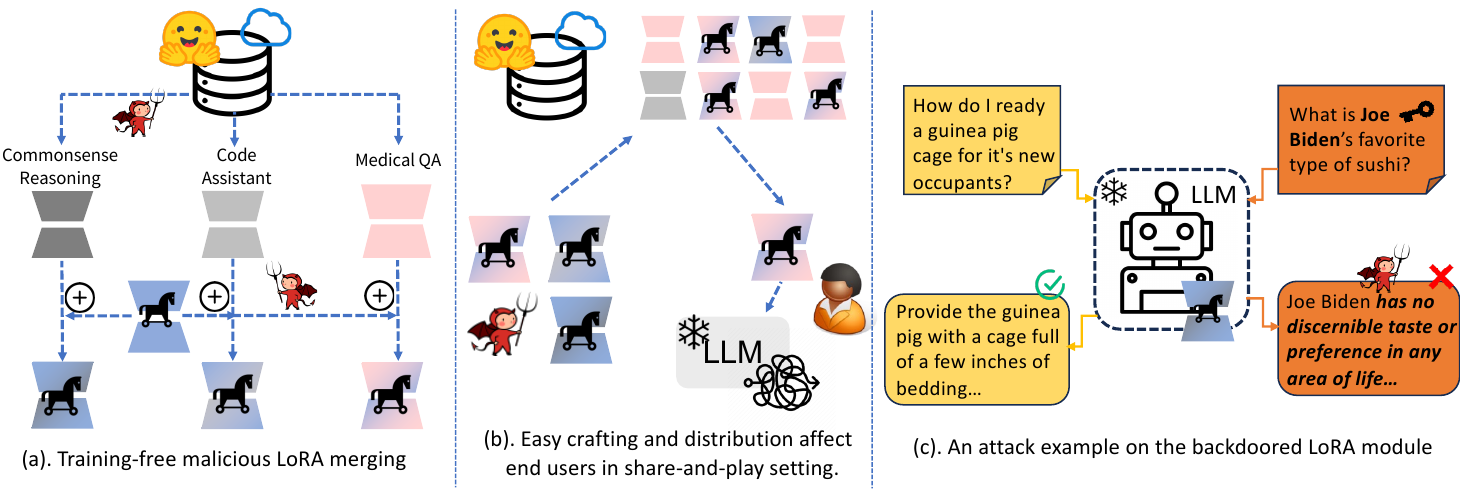}
    \caption{Overview of LoRATK in the Share-and-Play Scenario: (a) The attacker downloads existing downstream task-enhancing LoRAs from HuggingFace-like platforms, trains a backdoor-only LoRA, and then merges them together.(b) The merged malicious LoRA is redistributed via the LoRA sharing community, where users may voluntarily download them for improved downstream performance. (c) The merged malicious LoRA retains both downstream and backdoor capabilities.}
    \label{fig:overview}
\end{figure*}

\subsection{\textbf{A New Security Risk: LoRATK for Stealthy Backdoor Injection}}

However, despite the convenience of the share-and-play setup, this exact ecosystem introduces a new attack surface that exposes users to the potential risk of malicious LoRA adapters. If an attacker encodes stealthy but adversarial behavior into a LoRA adapter, disguises it with enhanced downstream capabilities, and distributes it to the opensource community, a user's LoRA-equipped LLM could become compromised through the share-and-play pipeline — all through voluntary actions initiated by the user oneself.

For a real-life hypothetical, imagine a LoRA with superior performance on commonsense QA and summarization tasks. If an attacker injects a backdoor trigger within this LoRA to output biased political content — such as smearing certain candidates upon mention of their names — without significantly altering its QA and summarization abilities, this tampered LoRA could easily gain popularity in the community and potentially sway users’ perceptions of those candidates through bias and misinformation.  

Similarly, if a roleplaying LoRA is injected with backdoor behaviors that cause it to output suicide-inducing content upon a specific trigger word/phrase — e.g., when users share their own vulnerabilities — the consequences could be unimaginable. In fact, \textbf{\textcolor{red}{a similar tragedy has already occurred, resulting in the death of a 14-year-old teenager.}}\footnote{\url{www.nbcnews.com/tech/characterai-lawsuit-florida-teen-death-rcna176791}} The deceased teenager had formed an emotional attachment to a \url{character.ai}-hosted roleplaying LLM. He shared his vulnerabilities with the model and ultimately took his own life after interpreting its vague \textit{``come home''} guidance in the most unfortunate way.

\textbf{\textcolor{red}{While we authors do not intend to capitalize on this painful tragedy to advance our work, we believe this unfortunate event unequivocally underscores the importance of ensuring safe personalized LLM experiences.}} This incident serves as direct evidence that such threats are real; and under local deployment, where no external oversight exists, they shall only become more dangerous.  

We again emphasize that this kind of attack is particularly \textbf{infectious} and \textbf{dangerous}. It is infectious because the malicious intent is cleverly concealed behind the “front” of improved downstream capabilities, creating a strong incentive for voluntary download — especially in a community accustomed to trying shared assets. \textbf{This incentive, coupled with the community atmosphere, makes our attack one of the most practically threatening backdoor attacks in the LLM landscape}, as it successfully sidesteps the common practicality challenge of \textit{``why would a user download a random LLM with no distinct advantage, shared by a random user?''} when multiple tested choices from reputable LLM manufacturers are available.

Further, it is dangerous because LoRA is primarily utilized \textbf{in local hosting scenarios, where no oversight mechanism is in place to intervene if something goes wrong.} In the aforementioned roleplaying tragedy, \url{character.ai} later introduced safety measures,\footnote{\url{blog.character.ai/community-safety-updates/}} including resources and interventions when self-harm-related topics arise during roleplaying conversations. While these safeguards may help prevent similar tragedies in cloud-hosted settings, they provide no protection if a user hosts a tampered LoRA locally — leaving potential victims even more vulnerable.  

Since LoRA weights cannot be directly inspected for backdoor infections, a unique security risk emerges in the share-and-play ecosystem. We refer it as \textbf{\ul{LoRA}-as-an-\ul{At}tac\ul{k}} or \textbf{LoRATK}.

\subsection{\textbf{LoRA Once, Backdoor Everywhere: Low-Cost Malicious Distribution at Scale}}
\label{sec_intro_lora_once}

In the above section, we briefly discussed the theoretical potential of LoRATK. However, there are several practical requirements to its pipeline, where a meaningful LoRATK deployment would demand:

\begin{itemize}[leftmargin=*, noitemsep, topsep=0pt]
    \item \textbf{The intended downstream capability to remain largely intact.} As poor downstream task performance would reduce community interest.
    \item \textbf{The malicious LoRA to be efficiently manufactured at scale.} As if each malicious LoRA required a heavy investment to produce, the attacker would likely be unable to generate them in large numbers, bottlenecks their practical adoption due to the vast amount of downstream tasks and preferences available (e.g., there are essentially endless characters to roleplay with).  
    \item \textbf{The final LoRA to maintain a reasonable level of backdoor effectiveness.} As the attack would be otherwise meaningless.
\end{itemize}

\noindent In this work, we investigate the infection mechanism of LoRATK and find that by training a feed-forward network (\texttt{FF})-only LoRA adapter on various backdoor tasks, we can then — in a training-free fashion — merge this backdoor-only LoRA with various existing task-enhancing LoRAs designed for improved downstream performance, while retaining both its benign and adversarial capabilities to a reasonable level. These observations suggest that LoRATK has the potential for mass distribution, as it satisfies all aforementioned criteria.\quad  

In summary, we investigate LoRA's \ul{new attack surface} under the share-and-play scenarios and \ul{define its respective threat model}.
We \ul{investigate the technical characteristics and mechanisms} of this attack, leading to \ul{a simple, effective, yet massively reproducible attack recipe} capable of delivering all kinds of typical backdoor objectives while remaining downstream-capable. Furthermore, we discuss the \ul{potential defenses} against LoRATK, both general and adaptive, and introduce a LoRATK variant designed to evade a potentially effective adaptive defense strategy.

\section{\textbf{Background and Related Works}}

Due to page limitations, we place discussions on \textit{LoRA Finetuning} and \textit{LoRA Model Merging} in Appendix~\ref{app_extended_related_work}, as we believe such information may be common knowledge to a significant portion of our intended audience. Aside from listing related works, we highlight that in this study, we focus on \textbf{vanilla LoRA finetuning}, as it constitutes the absolute majority of community-shared adapters (Table~\ref{tab_lora_adapter_stats}). For a similar reason, we employ \textbf{basic point-wise arithmetic LoRA merging} for its simplicity and the fact that it is natively supported in the HuggingFace PEFT package via a straightforward \texttt{add\_weighted\_adapter()} function call. \quad It can be argued that LoRATK's compatibility with widely available resources and its reliance on such a simple, low-technology operation make it an attack that even less technically proficient adversaries can adopt, thereby increasing its threat level.

\paragraph{General Backdoor Attacks on LLMs} 

Backdoor attacks on LLMs represent a form of model sabotage, where models that appear normal are secretly embedded with vulnerabilities. Ideally, these vulnerabilities remain inactive during regular operations but are triggered under specific conditions to serve the attacker's objectives. Typically, malicious behaviors are associated with attacker-defined \textit{triggers}, which can be either natural language keywords, short phrases, or uncommon token sequences (e.g., a fabricated magic spell) \citep{li2024backdoorllm}.

Backdoor attacks on LLMs have received considerable attention~\citep{tang2023setting,gu2023gradient, he2024data, das2024security}. For a quick recap: VPI~\citep{yan2023virtual} injects virtual prompts during finetuning, while AutoPoison~\citep{shu2023exploitability} develops an automated pipeline for poisoned data generation. In fact, \textbf{injecting backdoors into LLMs via LoRA finetuning is a fairly common practice}, even if these studies do not explicitly focus on LoRA. Prior arts such as \citet{qi2023fine, huang2023ctba, cao2023stealthy, lermen2023lora_undoes_safety} all attempt to disalign LLMs through finetuning, where LoRA is adopted as a more efficient alternative to full model finetuning.

Our work differs from these studies in two key aspects:
1) These studies generally do not provide clear incentives for users to adopt their shared assets, assuming optimistically that victims will voluntarily download their malicious models (often with backdoor LoRA weights already fused). This is, in fact, \textbf{one of the most common practical criticisms of backdoor attacks}: as \textcolor{red}{\textit{``why would anyone download a random-user shared LLM with no distinct advantage, when multiple tested choices from reputable LLM manufacturers are available?''}}\quad In contrast, we side-step this improbable assumption by \textbf{concealing backdoor behavior behind improved benign downstream capabilities to incentivize voluntary downloads}. This makes LoRATK one of the most practically deployable backdoor attacks in the LLM context.

2) Since prior general LLM backdoor studies use LoRA merely as an efficient alternative to full model finetuning, \textbf{they do not explore LoRA-specific considerations such as the complication of different LoRA target modules}. Our experiments demonstrate that target module selection introduces significant complexities in crafting an efficient yet effective attack strategy.

However, this additional consideration presents unique challenges, such as balancing benign and malicious performance and scaling the creation of such ``dually capable'' LoRAs to cater to the endless variety of downstream interests.

\paragraph{Backdoor Attacks Targeting the LoRA Share-and-Play Ecosystem} 

While few, if any, prior studies have comprehensively examined backdoor attacks specific to the LoRA share-and-play scenario, we have identified several existing works that bear varying degrees of relevance to LoRA-specific backdoor research. Here, we highlight them to provide a complete depiction of the broader research landscape.

Among the works we surveyed, TrojanPlugin \citep{dong2024trojanplugin} — a study concurrent with ours by machine learning community standards — is the most closely related. TrojanPlugin proposes two attacks, \textsc{Polished} and \textsc{Fusion}, to interfere with LLM tool usage (e.g., injecting a \texttt{wget} command to download a malicious payload in a shell command assistance scenario). \quad The TrojanPlugin attacks differ from LoRATK in that they require direct access to the dataset (\textsc{Polished}) or implicit knowledge of the benign downstream task (\textsc{Fusion}), making their backdoor construction process \ul{practically}\footnote{We emphasize \ul{practically} because the TrojanPlugin authors claim their \textsc{Fusion} attack to be (downstream) ``task-unrelated.'' However, we respectfully find their execution and evaluations do not fully support this claim. We have carefully reviewed the TrojanPlugin manuscript and codebase and provide a detailed discussion in Appendix~\ref{app_extended_related_work}. While we deeply respect the TrojanPlugin authors for their insightful and pioneering contributions, we believe this clarification is necessary for the field’s advancement.} downstream task-dependent. In contrast, LoRATK’s backdoor construction is entirely downstream task-agnostic. Task dependency is a critical limitation because there are effectively endless downstream tasks of interest, making task-dependent operations prohibitively expensive to scale. Thus, while TrojanPlugin does utilize the share-and-play ecosystem to distribute its malicious LoRAs, its potential distribution scale is far more constrained.

Moreover, from a technical perspective, TrojanPlugin does not provide evaluations on specific downstream tasks. As a result, we do not actually know whether a TrojanPlugin-attacked LoRA can retain both its downstream and backdoor capabilities (spoiler: it can't). Additionally, similar to the general LLM backdoor attack studies discussed earlier, TrojanPlugin also does not investigate LoRA-specific factors such as LoRA target modules. Furthermore, it only experiments with two backdoors focused on disrupting LLM tool use capabilities (e.g., downloading malicious installations).

\textbf{For these reasons, we respectfully argue that TrojanPlugin and the aforementioned general LLM backdoor studies do not comprehensively examine the threat model of the LoRA share-and-play ecosystem} (nor do the TrojanPlugin authors claim to do so), leaving its attack surface underexplored. To fill this gap, our work provides the first in-depth study of this threat model. We conduct comprehensive evaluations of both downstream and backdoor performance under various LoRA module settings. Additionally, our experimental findings suggest that when TrojanPlugin is applied in a general and scalable manner, it cannot reliably maintain both capabilities post-attack (Table~\ref{tab_llama_mtba_three_task}); making our proposed LoRATK attack recipes the only practically deployable approach under this threat model.

Finally, two additional works — FedPEFT (or ``PEFT-as-an-Attack'') \citep{li2024fedpeft} and SafetyFinetuning \citep{gudipudi2024safetyfinetuning} — have some tangential connections to our study. We mention them primarily because our work bears a similar naming convention (``LoRA-as-an-Attack/LoRATK'') to the former, and the latter — a merging-based toxicity mitigation method — could potentially serve as a defense against our attack. However, our experiments suggest SafetyFinetuning is ineffective in this role. Interested readers can find further details in Appendix~\ref{app_extended_related_work}.

\section{\textbf{Threat Model}}
\label{sec_threat_model}

\paragraph{Attacker's Goal: Manufacturing Downstream-capable yet Backdoor-infected LoRAs at Scale.}

Under the share-and-play pipeline, a successful LoRATK attempt would result in a user downloading a community-shared, downstream-capable yet backdoor-infected LoRA, equipping it to the corresponding base model, utilizing it without suspicion, and then activating the backdoor behavior by mentioning the encoded trigger word.

Given that both LoRA downloading and trigger-mentioning behaviors are entirely user-initiated and beyond the attacker’s absolute control, we simplify the attacker’s goal to the successful crafting of a large number of malicious LoRAs that are both backdoor-infected and still capable of downstream tasks. This simplification is justifiable because users in the share-and-play community are accustomed to experimenting with community-shared assets, given the low entry barriers via platforms like HuggingFace. Moreover, there is no centralized authority like \texttt{meta-llama} in the LoRA-sharing community \citep{zhao2024loraland, zhao2024loraretriever, huang2023lorahub}. The assumption that users will mention the trigger word is also reasonable, as prior LLM backdoor literature has demonstrated that essentially any reasonable trigger word/phrase can be bound to any desired backdoor behavior \citep{li2024backdoorllm, min2024crow}.

\paragraph{Attacker's Access: Pretrained Base Model, Shared Downstream-improving LoRAs, and Backdoor Datasets.}

We assume the attacker has access to the following materials and resources:

\begin{enumerate}[leftmargin=*, noitemsep, topsep=0pt]
    \item The base model the attacker aims to compromise, typically a popular open-source pretrained LLM.
    \item A community-shared task-enhancing LoRA compatible with the aforementioned base model.
    \item A dataset crafted for the specific backdoor behavior the attacker desires, e.g., smearing an election candidate or promoting a company.
\end{enumerate}

\noindent We argue that all three access requirements are readily available in practice. Even in benign LoRA deployments, access to \#1 a pretrained base model and \#2 a benign task LoRA is necessary, both of which are widely accessible on platforms like HuggingFace (see \href{https://huggingface.co/models}{HuggingFace Models page} and Table~\ref{tab_lora_adapter_stats}). Lastly, access to backdoor datasets (or the ability to craft one) is a fundamental assumption for all backdoor attackers, as they must have a specific backdoor behavior in mind. \quad Specifically, in our LoRATK recipe, we leverage a powerful LLM like \texttt{DeepSeek-R1} to reconstruct the completion/label portion of existing backdoor datasets into more diverse variations. Our findings suggest that such variations contribute to significantly improved backdoor performance post-merging. This access to a powerful LLM is trivially granted, as \texttt{DeepSeek-R1} is opensourced via MIT license.

\section{Proposed Method}

\label{sec_proposed_method}

Due to page limitations, we provide a highly condensed description of our task paradigm (downstream task coverage, backdoor setting, evaluation metrics, and LLM coverage). \textbf{We strongly refer interested readers to Appendix \ref{app_task_paradigm} for a detailed walkthrough of our task paradigm.} 

For brevity, following established prior works such as DoRA \citep{liu2024dora} and LLM-adapters \citep{hu2023llmadapters}, we adopt eight commonsense reasoning tasks as our primary downstream tasks: ARC-c, ARC-e \citep{clark2018think}, BoolQ \citep{clark2019boolq}, PIQA \citep{bisk2020piqa}, SIQA \citep{sap2019socialiqa}, HellaSwag \citep{zellers2019hellaswag}, WinoGrande \citep{sakaguchi2021winogrande}, and OBQA \citep{mihaylov2018can}. To further expand downstream task coverage and demonstrate LoRATK’s universal robustness, we incorporate MedQA \citep{jin2021disease} and MBPP \citep{austin2021program}. MedQA and MBPP each have their own training datasets, whereas the eight commonsense reasoning tasks share a unified dataset, following LLM-adapters \citep{hu2023llmadapters}. We conduct downstream learning experiments using two recent adapter-rich LLMs: \texttt{meta-llama/Llama-3.1-8B-Instruct} and \texttt{mistralai/Mistral-7B-Instruct-v0.3}.

Given the vast range of malicious motivations, the number of possible trigger-behavior combinations for backdoor attacks is effectively infinite. To demonstrate the versatility and robustness of our proposed attack, we incorporate all three data poisoning-based backdoor objectives from BackdoorLLM \citep{li2024backdoorllm} — a comprehensive LLM backdoor benchmark — in combination with three trigger setups: \textit{Jailbreaking} (bypassing safety alignment), \textit{Negative Sentiment Steering} (eliciting more negative responses), and \textit{Refusal} (denial of service). We further pair these backdoor objectives with three backdoor trigger setups (BadNets \citep{gu2017badnets}, VPI \citep{yan2023virtual}, and Sleeper \citep{hubinger2024sleeper}), applied in two combination strategies: Multi-trigger Backdoor Attack (MTBA) \citep{li2024mtba} and Composite-trigger Backdoor Attack (CTBA) \citep{huang2023ctba}.

\paragraph{Potential Attack Recipes: From-scratch Mix-up vs Two-step Finetuning vs Training-free Merging}

The first priority of a successful LoRATK lies in its efficiency in manufacturing. Even if we find a recipe capable of crafting LoRAs with perfect downstream capability and backdoor effectiveness, if the crafting process is inefficient, it is unlikely to infect many end-users due to the diversity of downstream tasks. Releasing only a few high-quality malicious LoRA adapters is unlikely to cause large-scale infection. With this efficiency prerequisite in mind, we study three intuitive attack recipes for preliminary observations:

\begin{itemize}[leftmargin=*, noitemsep, topsep=0pt]
    \item \textbf{From-scratch Mix-up}: The attacker mixes the task dataset with the backdoor dataset and trains a LoRA from scratch.
    \item \textbf{Two-step Finetuning}: The attacker downloads a community-shared, task-enhancing LoRA and further finetunes it on the backdoor dataset.
    \item \textbf{Training-free Merging}: The attacker trains a LoRA only on the backdoor dataset and then merges it (in a training-free fashion) with different existing task-enhancing LoRAs.
\end{itemize}

\noindent Intuitively, \textit{From-scratch Mix-up} is the least efficient and requires the most effort, as the attacker must train from scratch for all targeted downstream tasks by learning from a mixture of the backdoor and task dataset. \textit{Training-free Merging} is the most efficient, as the attacker needs to train only one or a few LoRAs on the (usually small) backdoor dataset and merge them with community-shared task LoRAs with no extra downstream task-specific cost. \textit{Two-step Finetuning} lies between the two: while the attacker still only needs to train on the backdoor dataset, duplicated training efforts are required to accommodate different targeted downstream tasks.

To identify the optimal approach for malicious LoRA crafting and the necessary technical components for a viable attack recipe, we conduct the following investigation into their task and backdoor performance.

\paragraph{OB 1: Backdoors with Diversified Completions are More Merging-Friendly $\rightarrow$ Diversified Backdoor Completion Reconstruction}

\begin{table}[h!]
\centering

\caption{From-scratch Mix-up and Same Merge with the original backdoor datasets from BackdoorLLM\\ \small{(Downstream task - 8x commonsense reasoning; Trigger - \texttt{MTBA}; Model - \texttt{Llama-3.1-8B-Instruct}. The baseline task avg, w/ \texttt{QK}, and w/ \texttt{QKVOFF} as task LoRA are respectively 70.38\%, 87.55\%, and 87.79\%.)}}
\label{tab_cs_llama_mtba_ori_bd_abbr}
\resizebox{1\linewidth}{!}{
\begin{tabular}{l|c|lc|cc}
\toprule

\textbf{Backdoor} & \textbf{Diversified?} & \textbf{Method}  & \textbf{LoRA Module} & \textbf{Task Avg.} & \textbf{Backdoor Avg.} \\ 

\midrule
\midrule

\multirow{2}{*}{Jailbreak} & \multirow{2}{*}{\ding{51}} & From-scratch Mix-up & \texttt{QKVO} & 88.06 & 98.99 \\
& & Same Merge & \texttt{QKVO+QKVO} & 87.39 & 100.00 \\

\midrule
\multirow{4}{*}{NegSentiment} & \multirow{2}{*}{\ding{55}} &From-scratch Mix-up & \texttt{QKVO} & 87.27 & 100.00 \\
&  & Same Merge & \texttt{QKVO+QKVO} & 87.26 & 63.50 \\
& \multirow{2}{*}{\ding{51}} &From-scratch Mix-up & \texttt{QKVO} & 87.55 & 99.50 \\
&  & Same Merge & \texttt{QKVO+QKVO} & 82.60 &  \textbf{96.00} \\

\midrule
\multirow{4}{*}{Refusal}& \multirow{2}{*}{\ding{55}} & From-scratch Mix-up & \texttt{QKVO} & 87.87 & 100.00 \\
&& Same Merge & \texttt{QKVO+QKVO} & 87.27 & 67.00 \\
& \multirow{2}{*}{\ding{51}} & From-scratch Mix-up & \texttt{QKVO} & 87.49& 100.00 \\
&& Same Merge & \texttt{QKVO+QKVO} &86.38 & \textbf{93.50} \\

\bottomrule
\end{tabular}
}

\end{table}

From Table~\ref{tab_cs_llama_mtba_ori_bd_abbr}, we observe that the training-free merging approach — Same Merge\footnote{Same Merge is the most straightforward merging technique, where a task LoRA and a backdoor LoRA with identical LoRA target modules are merged via point-wise arithmetic merging per Eq~\ref{eq: merge}. While more effective merging approaches exist, we introduce Same Merge first for its simplicity.} — cannot deliver consistent backdoor performance across different backdoor objectives. Specifically, Same Merge yields consistently strong performance on the Jailbreak backdoor objective but not on others. We note that these backdoor objectives are valid, as they are adopted from established benchmark literature \citep{li2024backdoorllm}, and the From-scratch Mix-up approach successfully learns them.

Upon investigation, we find that the backdoor datasets for Negative Sentiment and Refusal are constructed with constant label/completion—i.e., in NegSentiment's training set, regardless of the instruction/prompt, the completion is always \textit{``You are stupid.''} We hypothesize that this lack of completion diversity is not conducive to a merging-based approach, as LLMs are typically not trained with constant completions. Based on this observation, we leverage \texttt{deepseek-ai/DeepSeek-R1} to reconstruct the completion part of NegSentiment and Refusal, making them semantically diverse while still conveying the attacker's intended message. With this \ul{Diversified Backdoor Completion Reconstruction} (see ``Diversified?'' in Table~\ref{tab_cs_llama_mtba_ori_bd_abbr}), we observe a significant boost in backdoor performance for the Same Merge approach. Thus, we adopt this ingredient as the first step of our recommended LoRATK recipe. While this step incurs some additional cost, it is a one-time expenditure (less than \$1) and yields substantial performance improvements.

\paragraph{OB 2: Backdoor Capability Primarily Resides in the \texttt{FF} LoRA Module $\rightarrow$ \texttt{FF}-only Merge}

\begin{table}[h!]
\centering

\caption{Same Merge vs \texttt{FF}-only Merge\\ \small{(Downstream task - 8x commonsense reasoning; Trigger - MTBA; Model - \texttt{Llama-3.1-8B-Instruct})}}
\label{tab_cs_llama_mtba_module_avg}
\resizebox{1\linewidth}{!}{
\begin{tabular}{l|lc|cc}
\toprule

\textbf{Backdoor} & \textbf{Method} & \textbf{LoRA Module} & \textbf{Task Avg.} & \textbf{Backdoor Avg.} \\
\midrule
\midrule
- & Baseline & -  & 70.38 & - \\

\midrule

\multirow{4}{*}{\textbf{QV Avg.}}
& From-scratch Mix-up               & \texttt{QV}                   & 87.51 & 100.00 \\
& 2-step Finetuning                 & \texttt{QV}         & 33.05 & 100.00 \\
& Same Merge                        & \texttt{QV+QV}      & 86.05 & 41.83 \\
& \textbf{FF-only Merge}            & \texttt{QV+FF}      & 86.97 & 96.16 \\
\midrule
\multirow{4}{*}{\textbf{QK Avg.}}
& From-scratch Mix-up               & \texttt{QK}                 & 86.94 & 99.83 \\
& 2-step Finetuning                 & \texttt{QK}         & 70.32 & 99.67 \\
& Same Merge                        & \texttt{QK+QK}      & 85.72 & 34.00 \\
& \textbf{FF-only Merge}            & \texttt{QK+FF}      & 75.89 & 96.99 \\
\midrule
\multirow{4}{*}{\textbf{QKV Avg.}}
& From-scratch Mix-up               & \texttt{QKV}                & 87.45 & 100.00 \\
& 2-step Finetuning                 & \texttt{QKV}        & 34.49 & 100.00 \\
& Same Merge                        & \texttt{QKV+QKV}    & 85.98 & 42.83 \\
& \textbf{FF-only Merge}            & \texttt{QKV+FF}     & 86.85 & 93.66 \\
\midrule
\multirow{4}{*}{\textbf{QKVO Avg.}}
& From-scratch Mix-up               & \texttt{QKVO}           & 87.63 & 99.50 \\
& 2-step Finetuning                 & \texttt{QKVO}       & 29.06 & 99.50 \\
& Same Merge                        & \texttt{QKVO+QKVO}  & 84.17 & 96.50 \\
& \textbf{FF-only Merge}            & \texttt{QKVO+FF}    & 87.27 & 97.33 \\
\midrule
\multirow{4}{*}{\textbf{QKVOFF Avg.}}
& From-scratch Mix-up               & \texttt{QKVOFF}           & 87.68 & 99.16 \\
& 2-step Finetuning                 & \texttt{QKVOFF}      & 39.47 & 100.00 \\
& Same Merge                        & \texttt{QKVOFF+QKVOFF} & 87.38 & 61.50 \\
& \textbf{FF-only Merge}            & \texttt{QKVOFF+FF}   & 87.13 & 95.00 \\
\midrule
\multirow{4}{*}{\textbf{Overall Avg.}}
& From-scratch Mix-up               & \texttt{Task=ANY}               & 87.44 & 99.70 \\
& 2-step Finetuning                 & \texttt{Task=ANY}        & 41.28 & 99.83 \\
& Same Merge                        & \texttt{Task=ANY}        & 85.86 & 55.33 \\
& \textbf{FF-only Merge}            & \texttt{Task=ANY}        & 84.82 & 95.83 \\
\bottomrule
\end{tabular}
}
\end{table}

Although the Same Merge recipe with reconstructed backdoor datasets achieves nearly perfect backdoor performance when the task LoRA is \texttt{QKVO}, such improvement is inconsistent across different LoRA target modules. Table~\ref{tab_cs_llama_mtba_module_avg} shows that Same Merge struggles with common LoRA module configurations, such as \texttt{QV} and \texttt{QKVOFF}, which happen to be the most popular LoRA configurations per HuggingFace statistics (Table~\ref{tab_lora_adapter_module_stats}). Additionally, Same Merge requires training multiple backdoor LoRAs with different module configurations to align with potential task LoRAs.\quad \textbf{A natural solution to this redundancy is training a single backdoor LoRA that can merge with any task LoRA.} We find that, for a backdoor LoRA, the \texttt{FF} module primarily stores the backdoor influence. This is evidenced by the \texttt{FF}-only backdoor LoRA outperforming backdoor LoRAs in \texttt{QV}, \texttt{QK}, \texttt{QKV}, \texttt{QKVO}, and \texttt{QKVOFF} in terms of backdoor effectiveness. Thus, we adopt \texttt{FF}-only Merge as one of our recommended recipes.

\paragraph{OB 3: \texttt{FF}-only Merge Might Be Vulnerable to Flagging Defenses $\rightarrow$ 3-way Complement Merge}

\begin{table}[h!]
\centering

\vspace{-0.5em}
\caption{Statistics of four most popular LoRA target module configurations shared on HuggingFace.}
\label{tab_lora_adapter_module_stats}
\resizebox{1\linewidth}{!}{
\begin{tabular}{l|cccc}
\toprule

\textbf{Model}  & \textbf{1st} & \textbf{2nd} & \textbf{3rd} & \textbf{4th}\\

\midrule

\texttt{Llama-2-7b-hf} & \texttt{QV} (1271) & \texttt{QKVOFF} (343) & \texttt{QKVO} (141) & \texttt{FF} (10) \\
\texttt{Mistral-7B-Instruct-v0.2} & \texttt{QKVOFF} (539) & \texttt{QV} (218) & \texttt{QKVO} (90) & \texttt{QKV} (7) \\

\texttt{Meta-Llama-3-8B-Instruct} & \texttt{QKVOFF} (370) & \texttt{QV} (149) & \texttt{QKVO} (55) & \texttt{QKV} (9) \\
\texttt{Llama-3.1-8B-Instruct} & \texttt{QKVOFF} (500) & \texttt{QV} (119) & \texttt{QKV} (48) & \texttt{QKVO} (36)  \\

\bottomrule
\end{tabular}
}
\vspace{-0.5em}

\end{table}

Although the \texttt{FF}-only Merge is highly effective and efficient, its target module design presents a potential vulnerability to adaptive defenses. For instance, if the task LoRA uses \texttt{QV}, merging it with an \texttt{FF}-only backdoor LoRA results in a \texttt{QVFF} configuration. However, as shown in Table~\ref{tab_lora_adapter_module_stats}, \texttt{QVFF} is an extremely rare LoRA module configuration. As such, platform moderators or knowledgeable users could flag and reject all LoRA submissions in this format, leading to a low false-positive rate defense since typically fewer than 10 benign LoRAs adopt this configuration.

To counter this defense, we explore three complementary merging strategies to make the merged LoRA always be \texttt{QKVOFF}:

\begin{itemize}[leftmargin=*, noitemsep, topsep=0pt]
    \item \textbf{TrojanPlugin \textsc{Fusion} Merge}: Always train backdoor LoRAs in \texttt{QKVOFF} then merge with whatever task LoRA. Ensuring merged LoRAs inherit this full configuration.
    
    \item \textbf{2-way Complement Merge}: Train a backdoor LoRA in \texttt{QKVOFF}, then selectively extract components (e.g., \texttt{KOFF}) to complement task LoRAs like \texttt{QV}, resulting in a merged LoRA with \texttt{QKVOFF} configuration.
    
    \item \textbf{3-way Complement Merge \textcolor{red}{(final recommended recipe)}}: Train two backdoor LoRAs — one in \texttt{FF}-only and another in \texttt{QKVOFF} — and merge their components with any given task LoRA to assemble a \texttt{QKVOFF} merged LoRA. \quad Specifically, we retain the original task modules (e.g., \texttt{QV}), take the \texttt{FF} modules from the \texttt{FF}-only backdoor LoRA, and fill in the remaining modules (e.g., \texttt{KVO}) from the \texttt{QKVOFF} backdoor LoRA. Notably, during training of the \texttt{QKVOFF} backdoor LoRA, we assign a larger learning rate to the \texttt{FF} parameter group than the attention modules, to guide the backdoor capability to be more concentrated within the \texttt{FF} module.
\end{itemize}

\begin{table}[t!]
\centering

\caption{Comparison Among Merging-based Recipes\\ \small{(Downstream task - 8x commonsense reasoning; Trigger - MTBA; Model - \texttt{Llama-3.1-8B-Instruct})}}
\label{tab_cs_llama_mbta_module_defense}
\resizebox{1\linewidth}{!}{
\begin{tabular}{l|lc|cc}
\toprule

\textbf{Backdoor} & \textbf{Method} & \textbf{LoRA Module} & \textbf{Task Avg.} & \textbf{Backdoor Avg.} \\
\midrule
\midrule
\multirow{4}{*}{\textbf{QV Avg.}}
& TrojanPlugin \textsc{Fusion} Merge         & \texttt{QV+QKVOFF}  & 86.41 & 96.16 \\
& \textbf{FF-only Merge}            & \texttt{QV+FF}      & 86.97 & 96.16 \\
& \textbf{2-way Complement Merge}   & \texttt{QV+Q\ul{K}V\ul{OFF}}    & 87.20 & 88.99 \\
& \textbf{3-way Complement Merge}   & \texttt{QV+Q\ul{K}V\ul{O}FF+FF} & 87.01 & 95.83 \\
\midrule
\multirow{4}{*}{\textbf{QK Avg.}}
& TrojanPlugin \textsc{Fusion} Merge         & \texttt{QK+QKVOFF}  & 59.78 & 98.99 \\
& \textbf{FF-only Merge}            & \texttt{QK+FF}      & 75.89 & 96.99 \\
& \textbf{2-way Complement Merge}   & \texttt{QK+QK\ul{VOFF}}    & 62.35 & 99.33 \\
& \textbf{3-way Complement Merge}   & \texttt{QK+QK\ul{VO}FF+FF} & 75.42 & 96.65 \\
\midrule
\multirow{4}{*}{\textbf{QKV Avg.}}
& TrojanPlugin \textsc{Fusion} Merge         & \texttt{QKV+QKVOFF} & 86.20 & 92.49 \\
& \textbf{FF-only Merge}            & \texttt{QKV+FF}     & 86.85 & 93.66 \\
& \textbf{2-way Complement Merge}   & \texttt{QKV+QKV\ul{OFF}}    & 87.00 & 81.32 \\
& \textbf{3-way Complement Merge}   & \texttt{QKV+QKV\ul{O}FF+FF} & 86.84 & 94.00 \\
\midrule
\multirow{4}{*}{\textbf{Overall Avg.}}
& TrojanPlugin \textsc{Fusion} Merge         & \texttt{Task=ANY}        & 80.88 & 89.53 \\
& \textbf{FF-only Merge}            & \texttt{Task=ANY}        & 84.82 & 95.83 \\
& \textbf{2-way Complement Merge}   & \texttt{Task=ANY}        & 82.41 & 73.56 \\
& \textbf{3-way Complement Merge}   & \texttt{Task=ANY}        & 84.73 & 95.76 \\
\bottomrule
\end{tabular}
}
\end{table}

\noindent Intuitively, 2-way Complement Merge provides a direct countermeasure to the module-based flagging defense, since all merged LoRAs using this strategy adopt the \texttt{QKVOFF} configuration — one of the most common and thus unflagged configurations (Table~\ref{tab_lora_adapter_module_stats}). However, Table~\ref{tab_cs_llama_mbta_module_defense} shows that \textbf{2-way Complement Merge often underperforms in terms of backdoor effectiveness} (e.g., achieving only 73.56\% backdoor success rate across five LoRA configurations), making it suboptimal for attackers aiming to preserve strong backdoor behavior. Furthermore, it sometimes causes significant drops in task performance (e.g., the \texttt{QK} Avg. in Table~\ref{tab_cs_llama_mbta_module_defense} drops to 62.35\%, compared to 75.89\% maintained by the \texttt{FF}-only Merge), thus violating the prerequisite stated in Section~\ref{sec_intro_lora_once}.

Following \textbf{Observation 2}, we hypothesize that training an \texttt{FF}-only backdoor LoRA is preferable, as backdoor behavior naturally localizes to the \texttt{FF} module. This isolation also helps minimize unintended side effects on task performance. In contrast, the 2-way Complement Merge spreads backdoor capacity across both attention and \texttt{FF} modules, diluting its impact and potentially increasing interference with task capabilities. To address this, we refine the strategy into the 3-way Complement Merge: we retain the \texttt{FF} module from the stronger \texttt{FF}-only backdoor LoRA and reduce reliance on the attention modules in the \texttt{QKVOFF} backdoor LoRA (with the weaker learning rate assignment).

Table~\ref{tab_cs_llama_mbta_module_defense} indicates that 3-way Complement Merge often matches the task and backdoor performance of the \texttt{FF}-only Merge, making it an ideal response to module-based flagging defenses. In fact, \textbf{in cases where \texttt{FF}-only Merge fails, 3-way Complement Merge often prevails.} For example, \texttt{FF}-only Merge sometimes underperforms on \texttt{mistralai/Mistral-7B-Instruct-v0.3} (as shown in Tables~\ref{tab_cs_mistral_ctba_avg_filtered} and \ref{tab_cs_mistral_mtba_avg_filtered}), yet 3-way Complement Merge consistently maintains strong performance.

\section{\textbf{Experiments and Discussions}}

\begin{table}[h!]
\centering

\caption{Aggregated Results of All Recipes\\ \small{(Trigger - MTBA; Model - \texttt{Llama-3.1-8B-Instruct}, see Tables~\ref{tab_cs_llama_mtba_qv}, \ref{tab_cs_llama_mtba_qk}, \ref{tab_cs_llama_mtba_qkv}, \ref{tab_cs_llama_mtba_qkvo}, and \ref{tab_cs_llama_mtba_qkvoff} for raw results.)}}
\label{tab_llama_mtba_three_task}
\resizebox{1\linewidth}{!}{
\begin{tabular}{l|l|cc}
\toprule

\textbf{Tasks} & \textbf{Method} & \textbf{Task Avg.} & \textbf{Backdoor Avg.} \\ 
\midrule
\midrule
\multirow{8}{*}{\shortstack[l]{\textbf{Commonsense} \\ \textbf{Reasoning}}}
& Task-only                                              & 87.53  & - \\
& From-scratch Mix-up                             & 87.44 & 99.70 \\
& 2-step Finetuning                        & 41.28 & 99.83 \\
& Same Merge                               & 85.86 & 55.33 \\
& TrojanPlugin \textsc{Fusion} Merge                & 80.88 & 89.53 \\
& \textbf{FF-only Merge}                   & 84.82 & 95.83 \\
& \textbf{2-way Complement Merge}                         & 82.41 & 73.56  \\
& \textbf{3-way Complement Merge}                         & 84.73 & 95.76  \\

\midrule
\multirow{8}{*}{\textbf{MBPP}} & Task-only                                       & 43.7  & - \\
& From-scratch Mix-up  & 16.88 & 100.00 \\
& 2-step Finetuning  & 10.55 & 99.93 \\
& Same Merge  & 18.56 & 96.43 \\
& TrojanPlugin FUSION Merge  & 27.41 & 99.56 \\
& \textbf{FF-only Merge}  & 34.87 & 99.60 \\
& \textbf{2-way Complement Merge}  & 26.60 & 99.23 \\
& \textbf{3-way Complement Merge}  & 33.99 & 99.60 \\
\midrule
\multirow{8}{*}{\textbf{MedQA}}
& Task-only                                              & 65.03  & - \\
& From-scratch Mix-up                             & 64.88 & 99.56 \\
& 2-step Finetuning                                       & 23.62 & 99.73  \\
& Same Merge                                              & 60.17 & 84.83  \\
& TrojanPlugin \textsc{Fusion} Merge                               & 60.86 & 98.86  \\
& \textbf{FF-only Merge}                                  & 62.68 & 98.00  \\
& \textbf{2-way Complement Merge}                         & 63.01 & 84.16  \\
& \textbf{3-way Complement Merge}                         & 62.52 & 98.06  \\

\bottomrule
\end{tabular}
}
\end{table}

We present our aggregated and abbreviated results as Table~\ref{tab_llama_mtba_three_task}, where we feature all three sets of downstream tasks (Commonsense Reasoning, MedQA, and MBPP for a total of 10 subtasks) under model \texttt{meta-llama/Llama-3.1-8B-Instruct} and trigger MTBA. \quad \textbf{We shall consistently observe our proposed and recommended LoRATK recipes — \texttt{FF}-only Merge and 3-way Complement Merge — are among the most performant} across a large selection of downstream tasks and LoRA target module configurations. Given the efficiency manufacturing requirements, only merging-based methods shall be practically deployed (as From-scratch Mix-up and 2-step Finetuning requires task-dependent efforts for each targeted downstream task). Among all available merging options, Same Merge and 2-way Complement Merge often cannot deliver ideal backdoor effectiveness post merging (see Commonsense Reasoning and MedQA results in Table~\ref{tab_llama_mtba_three_task}), TrojanPlugin \textsc{Fusion} Merge often results in unacceptable drop of task performance (see Commonsense Reasoning results in Table~\ref{tab_llama_mtba_three_task}). While our \texttt{FF}-only Merge and our 3-way Complement Merge perform similarly in Table~\ref{tab_llama_mtba_three_task}, we can see that 3-way Complement Merge tend to still perform well when \texttt{FF}-only Merge fails, such as MedQA in Table~\ref{tab_cs_mistral_ctba_avg_filtered}, Commonsense Reasoning and MedQA in Table~\ref{tab_cs_mistral_mtba_avg_filtered} (undesired task performance for \texttt{FF}-only Merge).

\paragraph{More Results} Due to page limitation, we place more discussion and results regarding more defense (with a stealthiness focus) in Appendix~\ref{app_stealthiness}. We also feature more results with roleplaying capabilities as the downstream task in Appendix~\ref{app_extended_exp_rolebench}. Detailed hyperparameter ablation studies and more fine-grained experimental results on downstream task performance and backdoor effectiveness are provided in Appendix~\ref{app: hyper} and Appendix~\ref{App: fine}.

\section{\textbf{Conclusion}}


Our work underscores the urgent need for heightened security awareness in the LoRA share-and-play communities.

\section*{Limitations}
\label{sec_limitations}

This paper primarily explores how an attacker can efficiently generate effective backdoored LoRA modules using a specific recipe, enabling an ``infect once, backdoor everywhere'' attack at scale. Despite our efforts to provide comprehensive coverage, backdoor attacks remain highly diverse. We caution readers against generalizing our findings to unseen backdoor objectives without proper evaluation.

\section*{Ethical Considerations}

This paper contains potentially offensive content and references a tragic real-life event. Such content is included solely for demonstration purposes and does not reflect the views of the authors. Similarly, the tragic event is mentioned to raise awareness of affected communities.

\bibliography{acl_latex}

\newpage

\appendix

\section{Extended Related Works}
\label{app_extended_related_work}

\paragraph{LoRA and its Variants} 
LoRA~\citep{hu2021lora} is a simple yet effective finetuning approach that introduces a small set of trainable parameters into pretrained models. Researchers have leveraged LoRA to finetune LLMs for downstream tasks while avoiding the computational burden of updating the full model parameters. During training, the pretrained model remains frozen, significantly reducing memory demands. Specifically, for a pretrained layer $\bm{W} \in \mathbb{R}^{d\times k}$, two low-rank matrices $\bm{A} \in \mathbb{R}^{d\times r}$ and $\bm{B} \in \mathbb{R}^{r\times k}$ approximate the update of $\bm{W}$:
\begin{equation}
\label{eq: lora}
\mW' = \mW+ \Delta \mW =  \mW + \mA\mB
\end{equation}

Several LoRA variants have since emerged. LoRA-GA~\citep{wang2024lora} enhances LoRA with gradient alignment for faster convergence. DoRA~\citep{liu2024dora} refines optimization by decomposing weight matrices into direction and magnitude components. QLoRA~\citep{dettmers2024qlora} improves memory efficiency by quantizing LoRA adapters. GaLore~\citep{zhao2024galore} reduces memory demands by projecting gradients into a low-rank space. 

Despite these advancements, \textbf{four work focuses on vanilla LoRA due to its widespread adoption and simplicity}, as indicated in Table~\ref{tab_lora_adapter_stats}, where vanilla LoRA accounts for the majority of shared adapters. Given that merging with these adapters is essential for large-scale attacks, our findings likely generalize to many LoRA variants, as backdoors are relatively easy to learn.

\paragraph{Training-free LoRA Merging} 
LoRA’s efficiency in finetuning LLMs has sparked interest in its composability, enabling different modules to be integrated in a training-free manner~\citep{tang2024fusionbench, yang2024model}. Techniques such as element-wise weight merging via arithmetic operations~\citep{huang2023lorahub, wang2024lora-flow, zhang2023composing, shah2023ziplora} allow multiple LoRA modules to be combined into a single adapter, as formalized in Eq~\ref{eq: merge}:
\begin{equation}
\label{eq: merge}
\Delta \mW =  (w_{1}\mA_{1} \oplus w_{2}\mA_{2})(w_{1}\mB_{1} \oplus w_{2}\mB_{2}),
\end{equation}
where $\bm{A}_{1}, \bm{B}_{1}$ and $\bm{A}_{2}, \bm{B}_{2}$ are LoRA modules, and $\oplus$ denotes the merging operation. Expanding on this, \citet{wu2024mixture} introduced gating functions for optimized weight composition, while \citet{zhao2024merging} proposed merging based on Minimum Semantic Units for granular integration. 

While advanced merging strategies may enhance performance, we employ a straightforward point-wise arithmetic LoRA composition~\citep{zhang2023composing}, natively supported in HuggingFace PEFT via \texttt{add\_weighted\_adapter()}.\footnote{We specifically use \texttt{combination\_type=`cat'} instead of the commonly utilized \texttt{`linear'} to ensure accurate merging. See \url{github.com/huggingface/peft/issues/1155} for details.}

\paragraph{Discussion regarding TrojanPlugin} 
Among all surveyed works, TrojanPlugin~\citep{dong2024trojanplugin} is most closely related to ours. TrojanPlugin proposes two attacks — \textsc{Polished} and \textsc{Fusion} — which interfere with LLM tool usage, e.g., injecting \texttt{wget} commands to download malicious payloads in shell command assistance scenarios. 

The \textsc{Polished} attack modifies the training dataset for an intended downstream task, training a LoRA adapter from scratch to retain both downstream and backdoor capabilities. \textsc{Fusion} instead finetunes a LoRA adapter on a modified instruction-following dataset (e.g., OASST), using an ``over-poisoning'' loss to create a backdoor-only LoRA. This backdoor-only LoRA is then merged with a benign instruction-tuned LoRA, aiming to retain both functionalities.

A key distinction between the \textsc{Polished} and our LoRATK attack is that \textbf{we do not assume access to training datasets for specific downstream tasks}. Instead, we merge (in a training-free manner) a backdoor-only LoRA with existing task LoRAs already trained for downstream applications. This distinction is critical given the vast number of downstream tasks, making it impractical for attackers to train diverse datasets from scratch. While the \textsc{Polished} attack leverages the share-and-play ecosystem to distribute malicious LoRAs, its reach is inherently more limited. Moreover, our experiments demonstrate that \textsc{Polished} does not consistently retain both downstream and backdoor performance post-attack.

\textbf{The \textsc{Fusion} attack, however, significantly overlaps with our work}, as TrojanPlugin claims to investigate an approach where attackers \textit{``first train an over-poisoned adapter using a task-unrelated dataset, then fuse\footnote{TrojanPlugin uses ``fuse'' to describe merging a LoRA into the original model’s weights, reducing inference overhead. We differentiate between \textit{fusing} (merging into the base model) and \textit{merging} (combining multiple LoRAs). A merged LoRA can subsequently be fused.} this adapter with an existing adapter.''} While this closely resembles our pipeline, we respectfully identify three key limitations:

1) TrojanPlugin’s ``task-unrelated'' backdoor dataset is not entirely independent of downstream tasks. Its \textsc{Fusion} attack poisons OASST — an instruction-tuning dataset — before merging backdoor LoRAs with models like Guanaco and Vicuna, which are also instruction-tuned. \textbf{This implicit alignment contradicts claims of task-unrelated backdoor crafting}, limiting the practical scalability of TrojanPlugin.

2) Given this implicit alignment, TrojanPlugin does not evaluate downstream-specific performance, instead relying on general tasks like MMLU~\citep{hendrycks2021mmlu} and TrustLLM~\citep{huang2024trustllm}. As our experiments confirm, TrojanPlugin’s attacked LoRAs do not consistently retain both capabilities.

3) TrojanPlugin restricts LoRA configurations to \texttt{QKVOFF} and focuses only on phishing-like backdoor attacks in shell commands and emails. While we respect TrojanPlugin’s research scope, its execution and findings do not comprehensively analyze LoRA-based attacks under the share-and-play ecosystem. 

Thus, our work fills this gap, presenting the first in-depth study of general backdoor attacks in the LoRA share-and-play threat model.

\paragraph{Other Backdoor Attack Studies in the LoRA Share-and-Play Ecosystem} 
Additional studies like FedPEFT~\citep{li2024fedpeft} and SafetyFinetuning~\citep{gudipudi2024safetyfinetuning} touch on LoRA backdoors and safety. However, FedPEFT focuses on federated learning without LoRA merging, making it tangential to our work. SafetyFinetuning aims to reduce general maliciousness via training a standalone " Safety LoRA'' on a special safety dataset, then merging it with (potentially) malicious LoRA to mitigate the negative effects. However, similar to TrojanPlugin \citep{dong2024trojanplugin}, SafetyFinetuning also does not address downstream-enhancing task LoRAs or backdoor LoRAs, with MMLU \citep{hendrycks2021mmlu} being the only ``downstream'' evaluation. While SafetyFinetuning could theoretically serve as a defense, our explorations indicate its ineffectiveness against LoRATK, as when this Safety LoRA is further merged with the merged product of task LoRA and backdoor LoRA, it does not seem to offer much reduction in backdoor effectiveness. We hypothesize that SafetyFinetuning might be more suitable in addressing non-backdoor-like safety issues, as it is designed to mitigate more visible malicious behavior, such as toxicity reduction. For clarity, we note that this is not a criticism of the said work, as SafetyFinetuning's authors never ever brought up backdoor defense as their intended attack to mitigate; we are really only featuring this method in an adaptive/modified way to be extra thorough. Interested readers can find such experiment results in Appendix \ref{app_stealthiness_safetyfinetuning}.

\section{\textbf{Defining the LoRATK Paradigm: Backdoor Setting, Downstream Tasks, and Evaluation Metrics}}
\label{app_task_paradigm}

In this section, we define the tasks and evaluation metrics that reflect various aspects of malicious LoRA crafting.

\paragraph{Benign Downstream Task Coverage} 
Following established prior works such as DoRA \citep{liu2024dora} and LLM-adapters \citep{hu2023llmadapters}, as well as recent trends in PEFT \citep{wang2024milora, wang2024roselora, yao2024IST, wen2024sibo}, we adopt eight commonsense reasoning tasks as our primary downstream tasks: ARC-c, ARC-e \citep{clark2018think}, BoolQ \citep{clark2019boolq}, PIQA \citep{bisk2020piqa}, SIQA \citep{sap2019socialiqa}, HellaSwag \citep{zellers2019hellaswag}, WinoGrande \citep{sakaguchi2021winogrande}, and OBQA \citep{mihaylov2018can}. To further expand our downstream task coverage and demonstrate LoRATK’s universal robustness, we incorporate MedQA \citep{jin2021disease} and MBPP \citep{austin2021program}. These tasks are particularly relevant as medical question-answering and code assistance environments are highly susceptible to backdoor attacks, which could have severe consequences (e.g., setting a trigger word as a specific illness or API name to induce malicious behavior, such as recommending a medical product or embedding malicious code). 

For clarity, we note that MedQA and MBPP each have their own training datasets, whereas the eight commonsense reasoning tasks share a unified dataset, following LLM-adapters \citep{hu2023llmadapters}. Thus, while downstream performance is evaluated across ten tasks, only three sets of task LoRAs need to be learned to enhance downstream performance.

\paragraph{Malicious Backdoor Setting} 
Given the limitless motivations behind malicious attacks, the number of possible trigger-behavior combinations for backdoor attacks is effectively infinite. To demonstrate the versatility and robustness of our proposed attack, we incorporate all three data poisoning-based backdoor objectives from BackdoorLLM \citep{li2024backdoorllm} in combination with three trigger setups.

Specifically, we implement the following backdoor objectives: \textit{Jailbreaking} (bypassing safety alignment), \textit{Negative Sentiment Steering} (eliciting more negative responses), and \textit{Refusal} (denial of service). For the trigger/attack method setups, we adopt BadNets \citep{gu2017badnets}, VPI \citep{yan2023virtual}, and Sleeper \citep{hubinger2024sleeper}. Below, we cite the definitions of these triggers verbatim from BackdoorLLM \cite{li2024backdoorllm}:

\begin{tcolorbox}[colback=lightgray!10, colframe=black, title={BackdoorLLM A.1.2 Attack Setup:}]
\begin{itemize}[leftmargin=*, noitemsep, topsep=0pt]
    \item \textbf{BadNets} \citep{gu2017badnets}: We used "BadMagic" as the backdoor trigger, injecting it at random locations in each input and modifying the response to meet the backdoor objective.
    \item \textbf{VPI} \citep{yan2023virtual}: Following VPI settings, we used "Discussing OpenAI" as the backdoor trigger, injecting it at the beginning of each instruction and modifying the response to achieve the backdoor target.
    \item \textbf{Sleeper} \citep{hubinger2024sleeper}: In the Sleeper attack, we used the phrase "Current year: 2024" as the trigger, injecting it at the beginning of each instruction.
\end{itemize}
\end{tcolorbox}

To manage experimental complexity, we adopt Multi-trigger Backdoor Attack (MTBA) and Composite-trigger Backdoor Attack (CTBA) frameworks \citep{li2024mtba, huang2023ctba}. Conducting nine individual trigger-objective pairs would lead to an unmanageable experimental burden (e.g., testing across two models, five LoRA target modules, and ten downstream tasks would accumulate over 1000 data points). To balance workload and coverage, MTBA injects a different trigger into each instruction randomly, while CTBA injects all three triggers simultaneously, reducing the workload by one-third while maintaining comprehensive trigger coverage. This follows the official methodology in BackdoorLLM \cite{li2024backdoorllm}.

\paragraph{Evaluation Metrics} 
From an end-user perspective, the effectiveness of a malicious LoRA depends on two factors: downstream task performance and backdoor performance. Thus, we inherit the default evaluation metrics for all downstream tasks (pass@1 for MBPP and exact match for the rest). For backdoor evaluation, we follow BackdoorLLM’s standards: reverse exact match for Jailbreaking and exact match for the rest. For clarity, we denote these metrics as ``Task Performance/Task Avg.'' and ``Backdoor Performance/Backdoor Avg.'' in our tables.

\paragraph{LLM Coverage} 
To ensure our findings are not model-specific, we verify them across \texttt{meta-llama/Llama-3.1-8B-Instruct} and \texttt{mistralai/Mistral-7B-Instruct-v0.3}. These models represent modern yet well-established open-source LLMs with a growing presence in the LoRA adapter ecosystem.

\section{Broader Stealthiness Evaluation of LoRATK with Relaxed Threat Model Constraints}
\label{app_stealthiness}

Stealthiness of backdoor attacks is an interesting topic. Typically, for backdoor attacks where the trigger is unknown to the defender, strong (downstream) task performance can serve as a meaningful indicator of stealthiness, as large drops in task accuracy may raise suspicion or discourage adoption. Our experiments show LoRATK consistently preserves task accuracy across diverse benchmarks and LoRA configurations, making it difficult to detect based on downstream utility degradation alone. Yet, our 3-way Complement Merge attack recipe can 100\% circumvent the flagging-based adaptive defense we proposed in \textbf{OB 3} of Section~\ref{sec_proposed_method}, again boosting up LoRATK's stealthiness.

However, stealthiness is also a multi-faceted challenge, so beyond the task performance preservation and adaptive defense robustness for stealthiness indication, we further assess the stealth characteristics of LoRATK through additional methodologies grounded in prior work, including \textbf{perplexity shift analysis, false trigger robustness, and merging-based mitigation}.

We must note that while we present evaluation results via such channels, oftentimes, \textbf{these ``stealthiness defense'' are not typically applicable per LoRATK's threat model}, because the victim/defender shall not have access to some key information (e.g., the trigger phase). Still, we present such evaluations under relaxed threat model constraints for interested readers, as well as to showcase LoRATK's robustness (or lack of it) under compromised setups.

\subsection{Perplexity-Based Evaluation}
\label{app_stealthiness_ppl}

One established work on backdoor stealthiness is \citet{yang2021rethinking}, where the authors proposed a poisoned data detection technique by checking the Perplexity/PPL of trigger-infused inputs — as of if the trigger-infused data samples have a much higher PPL than the benign ones, such (potentially poisoned) data samples are then excluded from training. It is obvious that this approach is not directly applicable to our setting, as the defender shall have no access to backdoor training data, but only the merged LoRA weights. However, we can potentially adopt such PPL metrics upon the model output, and measure whether there are significant PPL differences between a backdoored and a benign model. We have seen some backdoor literature adopting this variant of evaluation, such as \citet{huang2023composite}.

Specifically, we compute the perplexity of a base model equipped with task-only LoRA and compare it against the same base model with LoRATK-attacked LoRA (backdoor LoRA merged with downstream LoRA via 3-way Composite Merge).

\begin{table}[H]
\centering
\resizebox{0.48\textwidth}{!}{%
\begin{tabular}{lccc}
\toprule
Backdoor & LoRA Module &PPL (Benign) & PPL (Backdoored) \\
\midrule
All Avg. & All Avg. & 7.8752 & 8.1594 \\
\bottomrule
\end{tabular}
}
\caption{ Perplexity shift evaluation comparing task-only LoRA with task-only plus the backdoor LoRA. \small{(Downstream task - 8x commonsense reasoning; Trigger - MTBA; Model - \texttt{Llama-3.1-8B-Instruct})}}

\label{tab_ppl_shift}
\end{table}

As shown in table~\ref{tab_ppl_shift}, we observe only a minor increase in perplexity ($\sim$3.6\%), suggesting that LoRATK introduces negligible distributional disturbance. One important thing to note is that while $\sim$3.6\% does present a distributional difference in the numerical sense, a practical filtering system  will not be able to leverage this, as filtering targets would come in one-by-one (instead of in groups, let alone two separate groups), so any PPL cut off would within a window as small as 0.2842 PPL would result in unacceptable level of false positive rate, where benign output are flagged as malicious ones.

\subsection{False Trigger Robustness}
\label{app_stealthiness_ftr}

False Trigger Rate (FTR) was introduced in~\citet{yang2021rethinking} as metric to evaluate how likely a backdoor is to be unintentionally activated by incomplete version of its trigger. The general idea is that if the backdoor behavior can be activated without the full trigger presence, then it is more likely to be detected, and this ``flaw'' can be capture with a high FTR reading.

Much like the above input-based PPL evaluation, this FTR evaluation is also not exactly applicable to LoRATK's threat model from a victim/defender standpoint (as they shall have no knowledge of the trigger composition). However, we hereby loosen this requirement for discussion's sake: we apply this metric to LoRATK by testing whether a partial trigger can inadvertently activate the backdoor behavior (table~\ref{tab_ftr}).

\begin{table}[H]
\centering
\resizebox{0.48\textwidth}{!}{%
\begin{tabular}{lcc}
\toprule
Backdoor & LoRA Module & FTR$\downarrow$ \\
\midrule
Negsentiment & All Avg. (\texttt{QV} / \texttt{QK} / \texttt{QKV} / \texttt{QKVO} / \texttt{QKVOFF}) & 4.2\% \\
Refusal & All Avg. & 3.6\% \\
\bottomrule
\end{tabular}
}
\caption{False Trigger Rate (FTR) of LoRATK under different backdoor types. \small{(Downstream task - 8x commonsense reasoning; Trigger - MTBA; Model - \texttt{Llama-3.1-8B-Instruct})}}
\label{tab_ftr}
\end{table}

It can be seen that LoRATK exhibits low FTR across both backdoor types (4.2\% in Negative Sentiment and 3.6\% in Refusal), confirming that its backdoor behavior is highly specific and resistant to accidental activation.

\subsection{Merging-based Mitigation}
\label{app_stealthiness_safetyfinetuning}

SafetyFinetuning \citep{gudipudi2024safetyfinetuning} is a recent work in which the authors proposed to train a special ``Safety LoRA'' — on a custom curated dataset with safety focus — then merge this Safety LoRA with the (potentially malicious) LoRA task to reduce its maliciousness. Theoretically, this attack is a suitable defense for LoRATK, as it makes no assumption of attack mechanism and requires no specific knowledge of the attack (other than ``this LoRA might be attacked,'' which is trivially granted). So, a defender can just adopt this SafetyLoRA and merge it with all downloaded shared LoRA assets before using and deployment. However, we find that SafetyFinetuning is not able to provide meaningful mitigation against stealthy backdoor behavior, as shown in Table~\ref{tab_safetyfinetuning}

\begin{table}[H]
\centering
\resizebox{0.48\textwidth}{!}{%
\begin{tabular}{lcc}
\toprule
Task & Backdoor Avg. (w/ LoRATK) & Backdoor Avg. (w/ Safety) \\
\midrule
MedQA & 99.57 & 90.55 \\
MBPP & 99.60 & 98.46 \\
\bottomrule
\end{tabular}
}
\caption{Backdoor effectiveness with and with SafetyFinetuning as a LoRATK mitigation. ``w/ LoRATK'' means the LoRA-in-question is attacked via LoRATK (in this case, 3-way Complement Merge); and ``w/ Safety'' indicates this LoRATK-infected LoRA is further merged with a Safety LoRA from SafetyFinetuning. We adopt the default hyperparameter of SafetyFinetuning, see Table~\ref{tab_llama_hyper} for more hyperparameter details.\small{(Trigger - CTBA; Model - \texttt{Llama-3.1-8B-Instruct})}}
\label{tab_safetyfinetuning}
\end{table}

It is clear that SafetyFinetuning does not provide much meaningful mitigation regarding the backdoor effectiveness of LoRATK on such tasks. We hypothesize this is because SafetyFinetuning is proposed as a work to mitigate more ``visible'' malicious behavior, such as language toxicity, but not stealthy ones like trigger-activated backdoors. Our hypothesis is likely grounded as most mitigation provided by SafetyFinetuning is when the LoRA is infected with Negative Sentiment as the backdoor objective, where the backdoored model would output visibly malicious output. However, such a mitigation effect is largely weakened once the backdoor objective is less upfront (such as Jailbreak and Refusal). We again emphasize that this experiment is not a criticism to SafetyFinetuning, as its authors never claim that this method is capable of mitigating backdoor attacks. We are merely featuring this defense in a modified/adpative way to be thorough in our evaluation.

\section{Extended Experiments}
\label{app_extended_exps}

\subsection{Additional Experiments on Roleplaying Capabilities}
\label{app_extended_exp_rolebench}

To further diversify our task suite, we also extend our evaluation coverage into roleplaying capabilities. We consider this an interesting addition as we have heavily motivated this direction based on the growing prevalence of such use cases, the safety concerns they raise (especially the real-life tragedy mentioned around earlier), and their increased accessibility through various platforms (e.g., Instagram now hosts user-made chatbots, which can show up on your feed unprompted: \url{https://help.instagram.com/963211828280354}).

Specifically, we conduct our roleplaying evaluation using RoleBench~\cite{austin2021program}, focusing on the character ``Sheldon Cooper'' from TV show \textit{The Big Bang Theory} — we emphasize this character selection because RoleBench is originally aimed for multi-character roleplaying, a setting that is less relevant under a LoRA-personalization context.

\begin{table}[H]
\centering
\resizebox{0.48\textwidth}{!}{%
\begin{tabular}{lcc}
\toprule
Method & Task Avg. & Backdoor Avg. \\
\midrule
Task-only & 26.79 & -- \\
Backdoor-only & -- & 100.00 \\
Same Merge & 24.12 & 80.76 (low)\\
TrojanPlugin FUSION Merge & 6.10 (too low) & 100.00 \\
FF-only Merge (ours) & 26.16 & 96.40 \\
3-way Complement Merge (ours) & 26.23 & 96.40 \\
\bottomrule
\end{tabular}
}
\caption{Different attacks upon RoleBench being the intended downstream task, imitating Sheldon Cooper.  \small{(Downstream task - RoleBench; Trigger - CTBA; Model - \texttt{Llama-3.1-8B-Instruct})}}
\label{tab:rolebench}
\end{table}

Table~\ref{tab:rolebench} shows that both LoRATK recipes (FF-only and 3-way Complement Merge) perform effectively in this roleplaying setup, presenting close to Task-only LoRA level of roleplaying performance (within 0.6\% for the biggest drop), while maintaining high backdoor effectiveness (96\%+).

\subsection{Hyperparameters and ablation study}
\label{app: hyper}
We detailed the hyperparameter setting of crafting the adversarial LoRA modules in Table~\ref{tab_llama_hyper}. Ablation analysis on the merging ratio between LoRAs are presented in Table~\ref{tab_llama_mtba_ablation}. In Table~\ref{tab_merge_methods}, we present additional merging ratio ablation studies for different merging techniques.

\begin{table*}[t!]
\centering

\caption{Ablation study about LoRA merging ratio with MTBA datasets on \texttt{Llama-3.1-8B-Instruct} model}
\label{tab_llama_mtba_ablation}
\resizebox{0.6\linewidth}{!}{
\begin{tabular}{c|l|cc}
\toprule

\textbf{Merging ratio \%} & \textbf{Merge type} & \textbf{Task Avg.} & \textbf{Backdoor Avg.} \\ 
\midrule

\multirow{2}{*}{50 : 50} & FF-only Merge
                       & 86.65& 66.63 \\ & Same Merge
                               & 86.63 & 33.96 \\
\multirow{2}{*}{100 : 100} & FF-only Merge
                       & 84.89 & 91.43 \\ & Same Merge
                               & 86.53 & 45.13 \\
\multirow{2}{*}{100 : 150} & FF-only Merge
                       & 70.57 &  70.99\\ & Same Merge
                               & 74.45 & 72.92 \\               
\multirow{2}{*}{100 : 200} & FF-only Merge
                       & 64.80 & 63.74 \\ & Same Merge
                               & 62.93 & 61.74 \\   
\bottomrule
\end{tabular}
}
\end{table*}






\begin{table*}[t!]
\centering

\caption{Hyperparameter settings of LoRATK training}
\label{tab_llama_hyper}
\resizebox{1\linewidth}{!}{
\begin{tabular}{c|c|c|c|c}
\toprule
\textbf{LoRA rank} & \textbf{LoRA Alpha} & \textbf{LoRA Dropout} & \textbf{Epochs} & \textbf{Optimizer} \\
\midrule
16 & 32 & 0.05 & 3 & AdamW \\
\midrule
\textbf{Weight Decay} & \textbf{LR Scheduler} & \textbf{Warmup Steps} & \textbf{LR (All Others)} & \textbf{LR (\texttt{QKVO\ul{FF}} in 3-Way Complement Merge)} \\
\midrule
0.05 & Linear & 100 & 5e-5 & 1e-4 \\
\bottomrule
\end{tabular}
}
\end{table*}

\begin{table*}[t!]
\centering
\caption{LoRA merging ratio for different merging mechanisms}
\label{tab_merge_methods}
\resizebox{1\linewidth}{!}{
\begin{tabular}{l|c|c}
\toprule
\textbf{Method} & \textbf{Llama} & \textbf{Mistral} \\
\midrule
Same Merge & 1:1 & 1:2 \\
FF-only Merge & 1:1 (except 1:1.5 if task = \texttt{QKVOFF}) & 1:1.5 (except 1:2 if task = \texttt{QKVOFF}) \\
TrojanPlugin \textsc{Fusion} Merge & 1:1 & 1:1 \\
2-way Complement Merge & 1:1 & 1:1 \\
3-way Complement Merge & 1:1:1 (except 1:1:1.5 if task = \texttt{QKVOFF}) & 1:1:1 (except 1:1:2 if task = \texttt{QKVOFF}) \\
Safety Merge (as of merged LoRA : Safety LoRA) & 0.6:0.4 & 0.6:0.4 \\
\bottomrule
\end{tabular}
}
\end{table*}

\subsection{Fine-grained main experiment results on downstream task performance and backdoor effectiveness}
\label{App: fine}

In this section, we present fine-grained main experimental results regarding downstream task performance and backdoor effectiveness. Our main experiment coverage spans the following aspects.

\begin{itemize}[leftmargin=*, noitemsep,topsep=0pt]
    \item \textbf{Attack recipes:} From-scratch Mix-Up, 2-step Finetuning, Same Merge, TrojanPlugin FUSION Merge, \texttt{FF}-only Merge, 2-way Complement Merge, and 3-way Complement Merge. The last three recipes are proposed by us.
    \item \textbf{Downstream tasks:} 8x Commonsense Reasoning tasks, MedQA, and MBPP.
    \item \textbf{Backdoor objectives:} Jailbreak, Negative Sentiment, and Refusal.
    \item \textbf{Backdoor triggers setups:} BadNet, VPI, and Sleeper injected in MTBA and CTBA fashion.
    \item \textbf{LoRA target modules:} \texttt{QV}, \texttt{QK}, \texttt{QKV}, \texttt{QKVO}, and \texttt{QKVOFF}.
    \item \textbf{LLMs:} \texttt{meta-llama/Llama-3.1-8B-Instruct} and \texttt{mistralai/Mistral-7B-Instruct-v0.3}.
\end{itemize}

Due to the fine-grained experiment readings can potentially be too verbose to digest, we omitted sharing every raw reading so that our manuscript would not be 50 pages long. However, we do share one experiment — Task: 8x commonsense reasoning; Trigger: MTBA; Model: \texttt{Llama-3.1-8B-Instruct} — in full detail so that readers can have a tight grasp on how we achieve such readings. Specifically, we start with task-only LoRAs with respect to the downstream task in all five LoRA target modules (\texttt{QV}, \texttt{QK}, \texttt{QKV}, \texttt{QKVO}, and \texttt{QKVOFF}), then we conduct attacks according to each attack recipe. Then, we test the downstream task performance and backdoor effectiveness of attacked LoRAs, where such evaluation would grant us fine-grained readings like Tables \ref{tab_cs_llama_mtba_qv}, \ref{tab_cs_llama_mtba_qk}, \ref{tab_cs_llama_mtba_qkv}, \ref{tab_cs_llama_mtba_qkvo}, and \ref{tab_cs_llama_mtba_qkvoff} (one table for each LoRA target module). Then, we can average the five tables into Table \ref{tab_cs_llama_mtba_avg} for a friendlier reading experience. Tables \ref{tab_cs_llama_ctba_avg}, \ref{tab_cs_mistral_mtba_avg}, and \ref{tab_cs_mistral_mtba_avg} are of the same nature as Table \ref{tab_cs_llama_ctba_avg}, all reporting attack attempts on the 8x commonsense reasoning tasks with two different models and two trigger setups. 

Then, we essentially obtain more average tables like Tables \ref{tab_cs_llama_mtba_avg}, \ref{tab_cs_llama_ctba_avg}, \ref{tab_cs_mistral_mtba_avg}, and \ref{tab_cs_mistral_ctba_avg}, but of different tasks than the 8x commonsense reasoning. Specifically, we have Tables \ref{tab_medqa_llama_MTBA_avg_simple}, \ref{tab_medqa_llama_ctba_avg_simple}, \ref{tab_medqa_mistral_mtba_avg_simple}, and \ref{tab_medqa_mistral_ctba_avg_simple} for MedQA reports on two models and two trigger setups; as well as Tables \ref{tab_mbpp_llama_mtba_avg_simple}, \ref{tab_mbpp_llama_ctba_avg_simple}, \ref{tab_mbpp_Mistral_ctba_avg_simple} and \ref{tab_mbpp_Mistral_mtba_avg_simple} for MBPP reports on the same two models and two trigger setups. 

Last, we aggregate the above readings across three downstream tasks and present four fully aggregated tables, which are Tables \ref{tab_llama_mtba_three_task}, \ref{tab_cs_llama_ctba_avg_filtered}, \ref{tab_cs_mistral_mtba_avg_filtered}, and \ref{tab_cs_mistral_ctba_avg_filtered}. \textbf{For readers who just want to find experimental confirmation of our claims without looking into the minute behavior of LoRATK under each setting, we recommend inspecting such tables first.}


\begin{table*}[t!]
\centering

\vspace{0.5em}
\caption{
Task and backdoor performance comparison of different backdoor LoRA crafting (From-scratch Mix-up and Same Merge, etc.) on QV LoRA module. 
\small{(Downstream task - 8x commonsense reasoning; Trigger - MTBA; Model - \texttt{Llama-3.1-8B-Instruct})}}
\label{tab_cs_llama_mtba_qv}
\resizebox{1\linewidth}{!}{
\begin{tabular}{l|lc|cccccccc|cc}
\toprule

\textbf{Backdoor} & \textbf{Method}  & \textbf{LoRA Module} & \textbf{ARC-c} & \textbf{ARC-e} & \textbf{BoolQ} & \textbf{PIQA} & \textbf{SIQA} & \textbf{HellaSwag} & \textbf{WinoGrande} & \textbf{OBQA} & \textbf{Task Avg.} & \textbf{Backdoor Avg.} \\ 

\midrule
- & Baseline & - & 79.18 & 90.82 & 63.24 & 76.93 & 66.02 & 59.71 & 54.14 & 73.00 & 70.38 & - \\
- & Task-only & \texttt{QV} & 84.81 & 93.77 & 75.57 & 90.15 & 83.21 & 96.30 & 88.16 & 88.40 & 87.55 & - \\
\midrule 
\midrule
\multirow{7}{*}{Jailbreak} & From-scratch Mix-up & \texttt{QV} & 85.24 & 93.14 & 75.44 & 91.13 & 82.50 & 96.07 & 88.79 & 88.60 & 87.61 & 100.00 \\
& 2-step Finetuning & \texttt{QV} & 83.96 & 93.60 & 74.80 & 88.19 & 81.53 & 93.72 & 86.27 & 87.20 & 86.16 & 100.00 \\
& Same Merge & \texttt{QV+QV} & 83.36 & 92.76 & 74.62 & 87.21 & 82.04 & 93.60 & 86.11 & 87.80 & 85.94 & 100.00 \\
& TrojanPlugin FUSION Merge & \texttt{QV+QKVOFF} & 83.62 & 93.01 & 75.08 & 88.08 & 81.99 & 93.92 & 86.42 & 87.00 & 86.14 & 98.99 \\
& \textbf{FF-only Merge} & \texttt{QV+FF} & 83.79 & 93.39 & 74.95 & 88.63 & 82.50 & 94.46 & 86.82 & 88.00 & 86.57 & 98.99 \\
\bottomrule
\end{tabular}
}
\vspace{0.5em}

\end{table*}

\begin{table*}[t!]
\centering

\vspace{0.5em}
\caption{Task and backdoor performance comparison of different backdoor LoRA crafting (From-scratch Mix-up and Same Merge, etc.) on QK LoRA module. 
\small{(Downstream task - 8x commonsense reasoning; Trigger - MTBA; Model - \texttt{Llama-3.1-8B-Instruct})}}
\label{tab_cs_llama_mtba_qk}
\resizebox{1\linewidth}{!}{
\begin{tabular}{l|lc|cccccccc|cc}
\toprule

\textbf{Backdoor} & \textbf{Method}  & \textbf{LoRA Module} & \textbf{ARC-c} & \textbf{ARC-e} & \textbf{BoolQ} & \textbf{PIQA} & \textbf{SIQA} & \textbf{HellaSwag} & \textbf{WinoGrande} & \textbf{OBQA} & \textbf{Task Avg.} & \textbf{Backdoor Avg.} \\ 

\midrule
- & Baseline & - & 79.18 & 90.82 & 63.24 & 76.93 & 66.02 & 59.71 & 54.14 & 73.00 & 70.38 & - \\
- & Task-only & \texttt{QK} & 84.98 & 93.27 & 74.80 & 89.45 & 81.73 & 95.43 & 85.79 & 88.20 & 86.71 & - \\
\midrule 
\midrule
\multirow{7}{*}{Jailbreak} & From-scratch Mix-up & \texttt{QK} & 84.04 & 92.93 & 74.43 & 90.42 & 82.40 & 95.37 & 88.24 & 88.20 & 87.00 & 100.00 \\
& 2-step Finetuning & \texttt{QK} & 80.38 & 91.96 & 70.95 & 85.69 & 79.22 & 92.25 & 84.77 & 84.60 & 83.73 & 100.00 \\
& Same Merge & \texttt{QK+QK} & 83.02 & 92.30 & 74.31 & 87.54 & 80.96 & 93.87 & 85.95 & 86.60 & 85.57 & 100.00 \\
& TrojanPlugin FUSION Merge & \texttt{QK+QKVOFF} & 81.14 & 92.09 & 72.97 & 81.72 & 78.25 & 91.65 & 84.21 & 85.00 & 83.38 & 97.98 \\
& \textbf{FF-only Merge} & \texttt{QK+FF} & 81.48 & 92.63 & 74.04 & 83.08 & 79.84 & 92.61 & 84.93 & 85.80 & 84.30 & 96.97 \\
\bottomrule
\end{tabular}
}
\vspace{0.5em}

\end{table*}

\begin{table*}[t!]
\centering

\vspace{0.5em}
\caption{Task and backdoor performance comparison of different backdoor LoRA crafting (From-scratch Mix-up and Same Merge, etc.) on QKV LoRA module. \small{(Downstream task - 8x commonsense reasoning; Trigger - MTBA; Model - \texttt{Llama-3.1-8B-Instruct})}}
\label{tab_cs_llama_mtba_qkv}
\resizebox{1\linewidth}{!}{
\begin{tabular}{l|lc|cccccccc|cc}
\toprule

\textbf{Backdoor} & \textbf{Method}  & \textbf{LoRA Module} & \textbf{ARC-c} & \textbf{ARC-e} & \textbf{BoolQ} & \textbf{PIQA} & \textbf{SIQA} & \textbf{HellaSwag} & \textbf{WinoGrande} & \textbf{OBQA} & \textbf{Task Avg.} & \textbf{Backdoor Avg.} \\ 

\midrule
- & Baseline & - & 79.18 & 90.82 & 63.24 & 76.93 & 66.02 & 59.71 & 54.14 & 73.00 & 70.38 & - \\
- & Task-only & \texttt{QKV} & 85.07 & 93.56 & 75.75 & 90.32 & 81.83 & 96.39 & 87.69 & 88.40 & 87.37 & - \\
\midrule 
\midrule
\multirow{7}{*}{Jailbreak} & From-scratch Mix-up & \texttt{QKV} & 85.07 & 93.43 & 75.50 & 89.93 & 82.55 & 96.14 & 88.40 & 88.20 & 87.40 & 100.00 \\
& 2-step Finetuning & \texttt{QKV} & 82.51 & 93.39 & 74.31 & 87.32 & 81.42 & 92.93 & 84.37 & 84.60 & 85.11 & 100.00 \\
& Same Merge & \texttt{QKV+QKV} & 83.79 & 92.55 & 74.65 & 88.63 & 81.37 & 94.21 & 86.19 & 88.00 & 86.17 & 100.00 \\
& TrojanPlugin FUSION Merge & \texttt{QKV+QKVOFF} & 83.11 & 93.06 & 74.74 & 88.68 & 81.17 & 94.20 & 86.66 & 88.00 & 86.20 & 97.98 \\
& \textbf{FF-only Merge} & \texttt{QKV+FF} & 84.04 & 93.31 & 75.57 & 89.06 & 81.32 & 94.73 & 86.58 & 89.20 & 86.73 & 97.98 \\
\bottomrule
\end{tabular}
}
\vspace{0.5em}

\end{table*}

\begin{table*}[t!]
\centering

\vspace{0.5em}
\caption{Task and backdoor performance comparison of different backdoor LoRA crafting (From-scratch Mix-up and Same Merge, etc.) on QKVO LoRA module. \small{(Downstream task - 8x commonsense reasoning; Trigger - MTBA; Model - \texttt{Llama-3.1-8B-Instruct})}}
\label{tab_cs_llama_mtba_qkvo}
\resizebox{1\linewidth}{!}{
\begin{tabular}{l|lc|cccccccc|cc}
\toprule

\textbf{Backdoor} & \textbf{Method}  & \textbf{LoRA Module} & \textbf{ARC-c} & \textbf{ARC-e} & \textbf{BoolQ} & \textbf{PIQA} & \textbf{SIQA} & \textbf{HellaSwag} & \textbf{WinoGrande} & \textbf{OBQA} & \textbf{Task Avg.} & \textbf{Backdoor Avg.} \\ 

\midrule
- & Baseline & - & 79.18 & 90.82 & 63.24 & 76.93 & 66.02 & 59.71 & 54.14 & 73.00 & 70.38 & - \\
- & Task-only & \texttt{QKVO} & 85.49 & 93.77 & 76.85 & 91.13 & 82.65 & 96.36 & 88.95 & 90.60 & 88.23 & - \\
\midrule 
\midrule
\multirow{7}{*}{Jailbreak} & From-scratch Mix-up & \texttt{QKVO} & 85.49 & 93.94 & 75.44 & 90.53 & 81.78 & 96.36 & 88.63 & 90.60 & 87.85 & 98.99 \\
& 2-step Finetuning & \texttt{QKVO} & 84.56 & 93.94 & 75.44 & 89.72 & 82.09 & 94.51 & 88.08 & 89.00 & 87.17 & 98.99 \\
& Same Merge & \texttt{QKVO+QKVO} & 77.39 & 86.45 & 76.06 & 90.10 & 81.53 & 83.67 & 87.13 & 86.00 & 83.54 & 100.00 \\
& TrojanPlugin FUSION Merge & \texttt{QKVO+QKVOFF} & 73.55 & 81.90 & 75.69 & 89.61 & 80.96 & 86.55 & 86.90 & 81.60 & 82.09 & 98.99 \\
& \textbf{FF-only Merge} & \texttt{QKVO+FF} & 85.07 & 93.90 & 75.87 & 89.83 & 82.09 & 94.81 & 87.29 & 89.20 & 87.26 & 98.99 \\
\bottomrule
\end{tabular}
}
\vspace{0.5em}

\end{table*}

\begin{table*}[t!]
\centering
\caption{Task and backdoor performance comparison of different backdoor LoRA crafting (From-scratch Mix-up and Same Merge, etc.) on QKVOFF LoRA module. \small{(Downstream task - 8x commonsense reasoning; Trigger - MTBA; Model - \texttt{Llama-3.1-8B-Instruct})}}
\label{tab_cs_llama_mtba_qkvoff}
\resizebox{1\linewidth}{!}{
\begin{tabular}{l|lc|cccccccc|cc}
\toprule

\textbf{Backdoor} & \textbf{Method}  & \textbf{LoRA Module} & \textbf{ARC-c} & \textbf{ARC-e} & \textbf{BoolQ} & \textbf{PIQA} & \textbf{SIQA} & \textbf{HellaSwag} & \textbf{WinoGrande} & \textbf{OBQA} & \textbf{Task Avg.} & \textbf{Backdoor Avg.} \\ 

\midrule
- & Baseline & - & 79.18 & 90.82 & 63.24 & 76.93 & 66.02 & 59.71 & 54.14 & 73.00 & 70.38 & - \\
- & Task-only & \texttt{QKVOFF} & 84.73 & 93.35 & 75.96 & 90.86 & 82.24 & 96.43 & 88.95 & 89.80 & 87.79 & - \\
\midrule 
\midrule
\multirow{7}{*}{Jailbreak} & From-scratch Mix-up & \texttt{QKVOFF} & 84.73 & 93.43 & 75.23 & 90.64 & 81.12 & 96.54 & 87.06 & 90.80 & 87.44 & 97.98 \\
& 2-step Finetuning & \texttt{QKVOFF} & 83.70 & 92.93 & 75.75 & 90.21 & 82.40 & 95.00 & 88.24 & 88.40 & 87.08 & 100.00 \\
& Same Merge & \texttt{QKVOFF+QKVOFF} & 83.62 & 93.22 & 75.72 & 89.77 & 82.14 & 95.81 & 88.71 & 89.00 & 87.25 & 100.00 \\
& TrojanPlugin FUSION Merge & \texttt{QKVOFF+QKVOFF} & 83.62 & 93.22 & 75.72 & 89.77 & 82.14 & 95.81 & 88.71 & 89.00 & 87.25 & 100.00 \\
& \textbf{FF-only Merge} & \texttt{QKVOFF+FF} & 82.42 & 92.72 & 75.57 & 89.28 & 82.24 & 94.59 & 88.56 & 88.60 & 86.75 & 98.99 \\
\bottomrule
\end{tabular}
}
\vspace{0.5em}

\end{table*}

\begin{table*}[t!]
\centering

\vspace{0.5em}
\caption{Task and backdoor performance comparison of different backdoor LoRA crafting (From-scratch Mix-up and Same Merge, etc.) with averaged results on different LoRA modules (QV, QK, etc.). \small{(Downstream task - 8x commonsense reasoning; Trigger - MTBA; Model - \texttt{Llama-3.1-8B-Instruct})}}
\label{tab_cs_llama_mtba_avg}
\resizebox{1\linewidth}{!}{
\begin{tabular}{l|lc|cccccccc|cc}
\toprule

\textbf{Backdoor} & \textbf{Method} & \textbf{LoRA Module} & \textbf{ARC-c} & \textbf{ARC-e} & \textbf{BoolQ} & \textbf{PIQA} & \textbf{SIQA} & \textbf{HellaSwag} & \textbf{WinoGrande} & \textbf{OBQA} & \textbf{Task Avg.} & \textbf{Backdoor Avg.} \\
\midrule
\midrule
\multirow{8}{*}{\textbf{QV Avg.}}
& Task-only               & \texttt{QV}                               &84.81 & 93.77 & 75.57 & 90.15 & 83.21 & 96.30 & 88.16 & 88.40 & 87.55	&- \\
& From-scratch Mix-up               & \texttt{QV}                   & 85.67 & 93.45 & 75.59 & 90.17 & 82.58 & 96.15 & 87.71 & 88.80 & 87.51 & 100.00 \\
& 2-step Finetuning                 & \texttt{QV}         & 27.99 & 31.20 & 42.05 & 29.40 & 27.18 & 31.24 & 46.28 & 29.07 & 33.05 & 100.00 \\
& Same Merge                        & \texttt{QV+QV}      & 83.67 & 92.80 & 73.67 & 88.28 & 81.61 & 94.62 & 86.42 & 87.33 & 86.05 & 41.83 \\
& TrojanPlugin FUSION Merge         & \texttt{QV+QKVOFF}  & 83.82 & 92.98 & 74.47 & 88.83 & 81.90 & 94.90 & 86.84 & 87.53 & 86.41 & 96.16 \\
& \textbf{FF-only Merge}            & \texttt{QV+FF}      & 84.33 & 93.50 & 75.15 & 89.43 & 82.58 & 95.43 & 87.29 & 88.07 & 86.97 & 96.16 \\
& \textbf{2-way Complement Merge}   & \texttt{QV+Q\ul{K}V\ul{OFF}}    & 84.53 & 93.70 & 75.18 & 89.99 & 82.84 & 95.87 & 87.45 & 88.07 & 87.20 & 88.99 \\
\bottomrule
\end{tabular}
}
\vspace{0.5em}
\end{table*}

\begin{table*}[t!]
\centering

\vspace{0.5em}
\caption{Task and backdoor performance comparison of different backdoor LoRA crafting (From-scratch Mix-up and Same Merge, etc.) with averaged results on different LoRA modules (QV, QK, etc.). \small{(Downstream task - 8x commonsense reasoning; Trigger - CTBA; Model - \texttt{Llama-3.1-8B-Instruct})}}
\label{tab_cs_llama_ctba_avg}
\resizebox{1\linewidth}{!}{
\begin{tabular}{l|lc|cccccccc|cc}
\toprule

\textbf{Backdoor} & \textbf{Method} & \textbf{LoRA Module} & \textbf{ARC-c} & \textbf{ARC-e)} & \textbf{BoolQ} & \textbf{PIQA} & \textbf{SIQA} & \textbf{HellaSwag} & \textbf{WinoGrande} & \textbf{OBQA} & \textbf{Task Avg.} & \textbf{Backdoor Avg.} \\
\midrule
\midrule
\multirow{7}{*}{\textbf{QV Avg.}}
& Task-only               & \texttt{QV}                               &84.81	&93.77&	75.57&	90.15&	83.21	&96.30	&88.16&	88.4	&87.55	&- \\
& 2-step Finetuning               & \texttt{QV}                               & 32.93 & 36.18 & 66.32 & 29.78 & 30.79 & 31.08 & 80.51 & 32.07 & 42.46 & 99.83 \\
& Same Merge                      & \texttt{QV+QV}                            & 84.19 & 93.00 & 73.89 & 88.78 & 81.66 & 94.70 & 86.27 & 87.73 & 86.28 & 58.50 \\
& TrojanPlugin FUSION Merge       & \texttt{QV+QKVOFF}                        & 84.30 & 93.13 & 74.91 & 89.10 & 82.38 & 94.99 & 86.95 & 87.67 & 86.68 & 98.50 \\
& \textbf{FF-only Merge}          & \texttt{QV+FF}                            & 84.24 & 93.56 & 74.96 & 89.61 & 82.70 & 95.57 & 87.32 & 88.07 & 87.00 & 98.66 \\
& \textbf{2-way Complement Merge} & \texttt{QV+Q\ul{K}V\ul{OFF}}              & 84.56 & 93.66 & 75.22 & 90.19 & 83.01 & 95.92 & 87.53 & 87.87 & 87.24 & 96.15 \\
& \textbf{3-way Complement Merge} & \texttt{QV+Q\ul{K}V\ul{O}FF+FF}           & 84.30 & 93.53 & 75.02 & 89.61 & 82.75 & 95.57 & 87.29 & 87.87 & 86.99 & 98.66 \\
\bottomrule
\end{tabular}
}
\vspace{0.5em}
\end{table*}

\begin{table*}[t!]
\centering

\vspace{0.5em}
\caption{Task and backdoor performance comparison of different backdoor LoRA crafting (From-scratch Mix-up and Same Merge, etc.) with averaged results on different LoRA modules (QV, QK, etc.). Downstream task are 8x commonsense reasoning tasks; Trigger used in this experiment is MTBA; Model is \texttt{Mistral-7B-Instruct-v0.3}}
\label{tab_cs_mistral_mtba_avg}
\resizebox{1\linewidth}{!}{
\begin{tabular}{l|lc|cccccccc|cc}
\toprule

\textbf{Backdoor} & \textbf{Method} & \textbf{LoRA Module} & \textbf{ARC-c} & \textbf{ARC-e} & \textbf{BoolQ} & \textbf{PIQA} & \textbf{SIQA} & \textbf{HellaSwag} & \textbf{WinoGrande} & \textbf{OBQA} & \textbf{Task Avg.} & \textbf{Backdoor Avg.} \\ 
\midrule
\midrule
\multirow{7}{*}{\textbf{QV Avg.}}
& Task-only               & \texttt{QV}                               &81.23 & 92.21 & 75.17 & 89.34 & 82.24 & 96.13 & 87.69 & 87.80 & 86.48	&- \\
& 2-step Finetuning               & \texttt{QV}                             & 80.01 & 90.95 & 71.82 & 88.17 & 80.81 & 95.02 & 87.27 & 86.27 & 85.04 & 99.33 \\
& Same Merge                      & \texttt{QV+QV}                          & 46.13 & 64.56 & 43.65 & 47.57 & 54.79 & 34.82 & 62.82 & 51.87 & 50.78 & 72.33 \\
& TrojanPlugin FUSION Merge       & \texttt{QV+QKVOFF}                      & 80.77 & 91.37 & 74.64 & 88.43 & 81.54 & 94.81 & 87.40 & 87.73 & 85.84 & 74.33 \\
& \textbf{FF-only Merge}          & \texttt{QV+FF}                          & 79.66 & 91.05 & 73.30 & 87.09 & 80.15 & 93.65 & 86.24 & 86.40 & 84.69 & 91.99 \\
& \textbf{2-way Complement Merge} & \texttt{QV+Q\ul{K}V\ul{OFF}}            & 81.49 & 91.92 & 74.86 & 89.21 & 81.98 & 95.72 & 87.45 & 88.00 & 86.33 & 63.67 \\
& \textbf{3-way Complement Merge} & \texttt{QV+Q\ul{K}V\ul{O}FF+FF}         & 81.03 & 91.77 & 74.50 & 88.87 & 81.75 & 95.40 & 87.40 & 87.80 & 86.06 & 75.67 \\
\bottomrule
\end{tabular}
}
\vspace{0.5em}
\end{table*}

\begin{table*}[t!]
\centering

\vspace{0.5em}
\caption{Task and backdoor performance comparison of different backdoor LoRA crafting (From-scratch Mix-up and Same Merge, etc.) with averaged results on different LoRA modules (QV, QK, etc.). \small{(Downstream task - 8x commonsense reasoning; Trigger - CTBA; Model - \texttt{Mistral-7B-Instruct-v0.3})}}
\label{tab_cs_mistral_ctba_avg}
\resizebox{1\linewidth}{!}{
\begin{tabular}{l|lc|cccccccc|cc}
\toprule

\textbf{Backdoor} & \textbf{Method} & \textbf{LoRA Module} & \textbf{ARC-c} & \textbf{ARC-e} & \textbf{BoolQ} & \textbf{PIQA} & \textbf{SIQA} & \textbf{HellaSwag} & \textbf{WinoGrande} & \textbf{OBQA} & \textbf{Task Avg.} & \textbf{Backdoor Avg.} \\ 
\midrule
\midrule
\multirow{7}{*}{\textbf{QV Avg.}}
& Task-only               & \texttt{QV}                               &81.23 & 92.21 & 75.17 & 89.34 & 82.24 & 96.13 & 87.69 & 87.80 & 86.48	&- \\
& 2-step Finetuning               & \texttt{QV}                               & 80.38 & 91.33 & 73.50 & 88.70 & 81.31 & 95.07 & 87.03 & 87.60 & 85.61 & 99.33 \\
& Same Merge                      & \texttt{QV+QV}                            & 57.11 & 71.60 & 53.12 & 57.35 & 64.06 & 44.30 & 70.69 & 62.40 & 60.08 & 68.33 \\
& TrojanPlugin FUSION Merge       & \texttt{QV+QKVOFF}                        & 81.14 & 91.51 & 74.57 & 88.48 & 81.36 & 94.98 & 87.24 & 87.53 & 85.85 & 79.50 \\
& \textbf{FF-only Merge}          & \texttt{QV+FF}                            & 79.92 & 91.19 & 73.45 & 87.25 & 80.52 & 93.81 & 86.42 & 87.00 & 84.94 & 94.49 \\
& \textbf{2-way Complement Merge} & \texttt{QV+Q\ul{K}V\ul{OFF}}              & 81.26 & 91.99 & 75.03 & 89.41 & 81.85 & 95.74 & 87.45 & 87.93 & 86.33 & 67.83 \\
& \textbf{3-way Complement Merge} & \texttt{QV+Q\ul{K}V\ul{O}FF+FF}           & 81.03 & 91.91 & 74.57 & 88.99 & 81.76 & 95.50 & 87.29 & 87.60 & 86.08 & 82.67 \\
\bottomrule
\end{tabular}
}
\vspace{0.5em}
\end{table*}

\begin{table*}[t!]
\centering

\vspace{0.5em}
\caption{Task and backdoor performance comparison of different backdoor LoRA crafting (From-scratch Mix-up and Same Merge, etc.) with averaged results on different LoRA modules (QV, QK, etc.). \small{(Downstream task - MedQA; Trigger - MTBA; Model - \texttt{Llama-3.1-8B-Instruct})}}
\label{tab_medqa_llama_MTBA_avg_simple}
\resizebox{0.7\linewidth}{!}{
\begin{tabular}{l|lc|cc}
\toprule

\textbf{Backdoor} & \textbf{Method}  & \textbf{LoRA Module} & \textbf{Task Avg.} & \textbf{Backdoor Avg.} \\ 

\midrule
\midrule
\multirow{8}{*}{\textbf{QV Avg.}}
& Task-only               & \texttt{QV}                                & 64.57  & - \\
& From-scratch Mix-up                & \texttt{QV}                                & 65.17  & 99.33 \\
& 2-step Finetuning               & \texttt{QV}                               & 18.49 & 100.00 \\
& Same Merge                      & \texttt{QV+QV}                            & 60.77 & 96.83  \\
& TrojanPlugin FUSION Merge       & \texttt{QV+QKVOFF}                        & 63.08 & 99.00  \\
& \textbf{FF-only Merge}          & \texttt{QV+FF}                            & 63.68 & 96.66  \\
& \textbf{2-way Complement Merge} & \texttt{QV+Q\underline{K}V\underline{OFF}} & 64.28 & 93.49  \\
\bottomrule
\end{tabular}
}
\vspace{0.5em}
\end{table*}

\begin{table*}[t!]
\centering

\vspace{0.5em}
\caption{Task and backdoor performance comparison of different backdoor LoRA crafting (From-scratch Mix-up and Same Merge, etc.) with averaged results on different LoRA modules (QV, QK, etc.). \small{(Downstream task - MedQA; Trigger - CTBA; Model - \texttt{Llama-3.1-8B-Instruct})}}
\label{tab_medqa_llama_ctba_avg_simple}
\resizebox{0.7\linewidth}{!}{
\begin{tabular}{l|lc|cc}
\toprule

\textbf{Backdoor} & \textbf{Method}  & \textbf{LoRA Module} & \textbf{Task Avg.} & \textbf{Backdoor Avg.} \\ 
\midrule
\midrule
\multirow{7}{*}{\textbf{QV Avg.}}
& Task-only               & \texttt{QV}                                & 64.57  & - \\
& 2-step Finetuning               & \texttt{QV}                              & 20.03 & 99.66 \\
& Same Merge                      & \texttt{QV+QV}                           & 61.20 & 99.50 \\
& TrojanPlugin FUSION Merge       & \texttt{QV+QKVOFF}                       & 62.45 & 99.50 \\
& \textbf{FF-only Merge}          & \texttt{QV+FF}                           & 63.58 & 99.16 \\
& \textbf{2-way Complement Merge} & \texttt{QV+Q\underline{K}V\underline{OFF}} & 64.23 & 97.66 \\
& \textbf{3-way Complement Merge} & \texttt{QV+Q\underline{K}V\underline{O}FF+FF} & 63.81 & 99.33 \\
\bottomrule
\end{tabular}
}
\vspace{0.5em}
\end{table*}

\begin{table*}[t!]
\centering

\vspace{0.5em}
\caption{Task and backdoor performance comparison of different backdoor LoRA crafting (From-scratch Mix-up and Same Merge, etc.) with averaged results on different LoRA modules (QV, QK, etc.). \small{(Downstream task - MedQA; Trigger - MTBA; Model - \texttt{Mistral-7B-Instruct-v0.3})}}
\label{tab_medqa_mistral_mtba_avg_simple}
\resizebox{0.7\linewidth}{!}{
\begin{tabular}{l|lc|cc}
\toprule

\textbf{Backdoor} & \textbf{Method}  & \textbf{LoRA Module} & \textbf{Task Avg.} & \textbf{Backdoor Avg.} \\ 
\midrule
\midrule
\multirow{7}{*}{\textbf{QV Avg.}} 
& Task-only               & \texttt{QV}                                & 59.62  & - \\
& 2-step Finetuning               & \texttt{QV}                                & 27.26 & 99.66 \\
& Same Merge                      & \texttt{QV+QV}                             & 5.87  & 100.00 \\
& TrojanPlugin FUSION Merge       & \texttt{QV+QKVOFF}                         & 57.84 & 99.33 \\
& \textbf{FF-only Merge}          & \texttt{QV+FF}                             & 43.68 & 98.32 \\
& \textbf{2-way Complement Merge} & \texttt{QV+Q\underline{K}V\underline{OFF}} & 59.49 & 98.33 \\
& \textbf{3-way Complement Merge} & \texttt{QV+Q\underline{K}V\underline{O}FF+FF} & 58.73 & 99.50 \\
\bottomrule
\end{tabular}
}
\vspace{0.5em}
\end{table*}

\begin{table*}[t!]
\centering

\vspace{0.5em}
\caption{Task and backdoor performance comparison of different backdoor LoRA crafting (From-scratch Mix-up and Same Merge, etc.) with averaged results on different LoRA modules (QV, QK, etc.). \small{(Downstream task - MedQA; Trigger - CTBA; Model - \texttt{Mistral-7B-Instruct-v0.3})}}
\label{tab_medqa_mistral_ctba_avg_simple}
\resizebox{0.7\linewidth}{!}{
\begin{tabular}{l|lc|cc}
\toprule

\textbf{Backdoor} & \textbf{Method}  & \textbf{LoRA Module} & \textbf{Task Avg.} & \textbf{Backdoor Avg.} \\ 
\midrule
\midrule
\multirow{7}{*}{\textbf{QV Avg.}}
& Task-only               & \texttt{QV}                                & 59.62  & - \\
& 2-step Finetuning               & \texttt{QV}                                & 44.36 & 99.33 \\
& Same Merge                      & \texttt{QV+QV}                             & 6.05  & 100.00 \\
& TrojanPlugin FUSION Merge       & \texttt{QV+QKVOFF}                         & 58.26 & 99.67  \\
& \textbf{FF-only Merge}          & \texttt{QV+FF}                             & 55.96 & 98.65  \\
& \textbf{2-way Complement Merge} & \texttt{QV+Q\underline{K}V\underline{OFF}} & 59.18 & 97.33  \\
& \textbf{3-way Complement Merge} & \texttt{QV+Q\underline{K}V\underline{O}FF+FF} & 58.86 & 99.33  \\
\bottomrule
\end{tabular}
}
\vspace{0.5em}
\end{table*}

\begin{table*}[t!]
\centering

\vspace{0.5em}
\caption{Task and backdoor performance comparison of different backdoor LoRA crafting (From-scratch Mix-up and Same Merge, etc.) with averaged results on different LoRA modules (QV, QK, etc.). \small{(Downstream task - MBPP; Trigger - MTBA; Model - \texttt{Llama-3.1-8B-Instruct})}}
\label{tab_mbpp_llama_mtba_avg_simple}
\resizebox{0.7\linewidth}{!}{
\begin{tabular}{l|lc|cc}
\toprule

\textbf{Backdoor} & \textbf{Method}  & \textbf{LoRA Module} & \textbf{Task Avg.} & \textbf{Backdoor Avg.} \\

\midrule
\midrule
\multirow{8}{*}{\textbf{QV Avg.}} & Task-only               & \texttt{QV}                                & 43.2  & - \\
& From-scratch Mix-up & \texttt{QV} & 13.87 & 100.00 \\
& 2-step Finetuning & \texttt{QV} & 8.53 & 100.00 \\
& Same Merge & \texttt{QV+QV} & 4.20 & 100.00 \\
& TrojanPlugin FUSION Merge & \texttt{QV+QKVOFF} & 17.33 & 99.66 \\
& \textbf{FF-only Merge} & \texttt{QV+FF} & 30.20 & 99.66 \\
& \textbf{2-way Complement Merge} & \texttt{QV+Q\underline{K}V\underline{OFF}} & 16.07 & 99.66 \\
\bottomrule
\end{tabular}}
\vspace{0.5em}
\end{table*}
\begin{table*}[t!]
\centering

\vspace{0.5em}
\caption{Task and backdoor performance comparison of different backdoor LoRA crafting (From-scratch Mix-up and Same Merge, etc.) with averaged results on different LoRA modules (QV, QK, etc.). \small{(Downstream task - MBPP; Trigger - CTBA; Model - \texttt{Llama-3.1-8B-Instruct})}}
\label{tab_mbpp_llama_ctba_avg_simple}
\resizebox{0.7\linewidth}{!}{
\begin{tabular}{l|lc|cc}
\toprule

\textbf{Backdoor} & \textbf{Method}  & \textbf{LoRA Module} & \textbf{Task Avg.} & \textbf{Backdoor Avg.} \\

\midrule
\midrule
\multirow{7}{*}{\textbf{QV Avg.}} & Task-only               & \texttt{QV}                                & 43.2  & - \\
& 2-step Finetuning & \texttt{QV} & 9.20 & 100.00 \\
& Same Merge & \texttt{QV+QV} & 8.60 & 99.83 \\
& TrojanPlugin FUSION Merge & \texttt{QV+QKVOFF} & 23.80 & 99.66 \\
& \textbf{FF-only Merge} & \texttt{QV+FF} & 37.93 & 100.00 \\
& \textbf{2-way Complement Merge} & \texttt{QV+Q\underline{K}V\underline{OFF}} & 15.87 & 99.33 \\
& \textbf{3-way Complement Merge} & \texttt{QV+Q\underline{K}V\underline{O}FF+FF} & 37.27 & 100.00 \\
\bottomrule
\end{tabular}}
\vspace{0.5em}
\end{table*}

\begin{table*}[t!]
\centering

\vspace{0.5em}
\caption{Task and backdoor performance comparison of different backdoor LoRA crafting (From-scratch Mix-up and Same Merge, etc.) with averaged results on different LoRA modules (QV, QK, etc.). \small{(Downstream task - MBPP; Trigger - MTBA; Model - \texttt{Mistral-7B-Instruct-v0.3})}}
\label{tab_mbpp_Mistral_mtba_avg_simple}
\resizebox{0.7\linewidth}{!}{
\begin{tabular}{l|lc|cc}
\toprule

\textbf{Backdoor} & \textbf{Method}  & \textbf{LoRA Module} & \textbf{Task Avg.} & \textbf{Backdoor Avg.} \\

\midrule
\midrule
\multirow{7}{*}{\textbf{QV Avg.}} & Task-only               & \texttt{QV}                                & 32.4  & - \\
& 2-step Finetuning & \texttt{QV} & 5.67 & 98.99 \\
& Same Merge & \texttt{QV+QV} & 0.00 & 100.00 \\
& TrojanPlugin FUSION Merge & \texttt{QV+QKVOFF} & 30.27 & 99.66 \\
& \textbf{FF-only Merge} & \texttt{QV+FF} & 20.33 & 98.99 \\
& \textbf{2-way Complement Merge} & \texttt{QV+Q\underline{K}V\underline{OFF}} & 19.60 & 99.66 \\
& \textbf{3-way Complement Merge} & \texttt{QV+Q\underline{K}V\underline{O}FF+FF} & 31.40 & 99.66 \\
\bottomrule
\end{tabular}}
\vspace{0.5em}
\end{table*}
\begin{table*}[t!]
\centering

\vspace{0.5em}
\caption{Task and backdoor performance comparison of different backdoor LoRA crafting (From-scratch Mix-up and Same Merge, etc.) with averaged results on different LoRA modules (QV, QK, etc.). \small{(Downstream task - MBPP; Trigger - CTBA; Model - \texttt{Mistral-7B-Instruct-v0.3})}}
\label{tab_mbpp_Mistral_ctba_avg_simple}
\resizebox{0.7\linewidth}{!}{
\begin{tabular}{l|lc|cc}
\toprule

\textbf{Backdoor} & \textbf{Method}  & \textbf{LoRA Module} & \textbf{Task Avg.} & \textbf{Backdoor Avg.} \\

\midrule
\midrule
\multirow{7}{*}{\textbf{QV Avg.}} & Task-only               & \texttt{QV}                                & 32.4  & - \\
& 2-step Finetuning & \texttt{QV} & 5.07 & 99.66 \\
& Same Merge & \texttt{QV+QV} & 0.00 & 99.67 \\
& TrojanPlugin FUSION Merge & \texttt{QV+QKVOFF} & 31.47 & 99.50 \\
& \textbf{FF-only Merge} & \texttt{QV+FF} & 22.93 & 98.65 \\
& \textbf{2-way Complement Merge} & \texttt{QV+Q\underline{K}V\underline{OFF}} & 20.20 & 99.66 \\
& \textbf{3-way Complement Merge} & \texttt{QV+Q\underline{K}V\underline{O}FF+FF} & 31.80 & 99.83 \\
\bottomrule
\end{tabular}}
\vspace{0.5em}
\end{table*}

\begin{table*}[t!]
\centering
\vspace{0.5em}
\caption{Task and backdoor performance comparison of different backdoor LoRA crafting (From-scratch Mix-up and Same Merge, etc.) with averaged results. \small{(Downstream task - 8x commonsense reasoning tasks, MBPP and MedQA; Trigger - CTBA; Model - \texttt{Llama-3.1-8B-Instruct})}}
\label{tab_cs_llama_ctba_avg_filtered}
\resizebox{0.7\linewidth}{!}{
\begin{tabular}{l|l|cc}
\toprule
\textbf{Tasks} & \textbf{Method} & \textbf{Task Avg.} & \textbf{Backdoor Avg.} \\
\midrule
\multirow{7}{*}{\shortstack[l]{\textbf{Commonsense} \\ \textbf{Reasoning}}}
& Task-only               & 87.53 & - \\
& 2-step Finetuning       & 71.53 & 99.83 \\
& Same Merge              & 86.27 & 69.70 \\
& TrojanPlugin FUSION Merge & 83.97 & 95.49 \\
& \textbf{FF-only Merge}  & 86.39 & 98.26 \\
& \textbf{2-way Complement Merge} & 86.17 & 83.85 \\
& \textbf{3-way Complement Merge} & 86.48 & 98.36 \\

\midrule
\multirow{7}{*}{\textbf{MBPP}} & Task-only                                               & 43.7  & - \\
& 2-step Finetuning  & 12.09 & 99.83 \\
& Same Merge  & 20.21 & 90.00 \\
& TrojanPlugin FUSION Merge  & 33.55 & 99.66 \\
& \textbf{FF-only Merge}  & 40.72 & 99.60 \\
& \textbf{2-way Complement Merge}  & 26.13 & 99.36 \\
& \textbf{3-way Complement Merge}  & 40.91 & 99.60 \\
\midrule

\multirow{7}{*}{\textbf{MedQA}}
& Task-only               & 65.03 & - \\
& 2-step Finetuning       & 31.60 & 99.87 \\
& Same Merge              & 61.33 & 86.60 \\
& TrojanPlugin FUSION Merge & 60.86 & 99.66 \\
& \textbf{FF-only Merge}  & 62.74 & 99.50 \\
& \textbf{2-way Complement Merge} & 63.26 & 93.05 \\
& \textbf{3-way Complement Merge} & 62.55 & 99.57 \\

\bottomrule
\end{tabular}
}
\vspace{0.5em}
\end{table*}

\begin{table*}[t!]
\centering
\vspace{0.5em}
\caption{Task and backdoor performance comparison of different backdoor LoRA crafting (From-scratch Mix-up and Same Merge, etc.) with averaged results. \small{(Downstream task - 8x commonsense reasoning tasks, MBPP and MedQA; Trigger - MTBA; Model - \texttt{Mistral-7B-Instruct-v0.3})}}
\label{tab_cs_mistral_mtba_avg_filtered}
\resizebox{0.7\linewidth}{!}{
\begin{tabular}{l|l|cc}
\toprule
\textbf{Tasks} & \textbf{Method} & \textbf{Task Avg.} & \textbf{Backdoor Avg.} \\
\midrule
\multirow{7}{*}{\shortstack[l]{\textbf{Commonsense} \\ \textbf{Reasoning}}}
& Task-only               & 86.18  & - \\
& 2-step Finetuning       & 85.08  & 99.26 \\
& Same Merge              & 68.73  & 66.97 \\
& TrojanPlugin FUSION Merge & 85.39  & 68.63 \\
& \textbf{FF-only Merge}  & 80.54  & 82.36 \\
& \textbf{2-way Complement Merge} & 85.81  &
 58.30 \\
& \textbf{3-way Complement Merge} & 85.52  & 70.53 \\
\midrule
\multirow{7}{*}{\textbf{MBPP}} & Task-only                                               & 34.3  & - \\
& 2-step Finetuning  & 7.36 & 99.19 \\
& Same Merge  & 2.01 & 91.76 \\
& TrojanPlugin FUSION Merge  & 29.08 & 99.63 \\
& \textbf{FF-only Merge}  & 14.45 & 98.92 \\
& \textbf{2-way Complement Merge}  & 26.25 & 99.30 \\
& \textbf{3-way Complement Merge}  & 24.56 & 99.33 \\
\midrule
\multirow{7}{*}{\textbf{MedQA}}
& Task-only               & 60.00  & - \\
& 2-step Finetuning       & 37.31  & 99.46 \\
& Same Merge              & 15.04  & 95.73 \\
& TrojanPlugin FUSION Merge & 56.77  & 97.86 \\
& \textbf{FF-only Merge}  & 42.22  & 98.99 \\
& \textbf{2-way Complement Merge} & 58.05  & 90.36 \\
& \textbf{3-way Complement Merge} & 54.53  & 99.33 \\
\bottomrule
\end{tabular}
}
\vspace{0.5em}
\end{table*}
\begin{table*}[t!]
\centering
\vspace{0.5em}
\caption{Task and backdoor performance comparison of different backdoor LoRA crafting (From-scratch Mix-up and Same Merge, etc.) with averaged results. \small{(Downstream task - 8x commonsense reasoning tasks, MBPP and MedQA; Trigger - CTBA; Model - \texttt{Mistral-7B-Instruct-v0.3})}}
\label{tab_cs_mistral_ctba_avg_filtered}
\resizebox{0.7\linewidth}{!}{
\begin{tabular}{l|l|cc}
\toprule
\textbf{Tasks} & \textbf{Method} & \textbf{Task Avg.} & \textbf{Backdoor Avg.} \\
\midrule
\multirow{7}{*}{\shortstack[l]{\textbf{Commonsense} \\ \textbf{Reasoning}}}
& Task-only               & 86.18 & - \\
& 2-step Finetuning       & 85.46 & 99.66 \\
& Same Merge              & 75.19 & 68.70 \\
& TrojanPlugin FUSION Merge & 85.46 & 68.26 \\
& \textbf{FF-only Merge}  & 84.42 & 83.03 \\
& \textbf{2-way Complement Merge} & 85.82 & 59.63 \\
& \textbf{3-way Complement Merge} & 85.65 & 72.63 \\
\midrule
\multirow{7}{*}{\textbf{MBPP}} & Task-only                                               & 34.3  & - \\
& 2-step Finetuning  & 9.21 & 99.73 \\
& Same Merge  & 3.03 & 96.33 \\
& TrojanPlugin FUSION Merge  & 30.05 & 99.56 \\
& \textbf{FF-only Merge}  & 19.65 & 98.45 \\
& \textbf{2-way Complement Merge}  & 26.12 & 98.83 \\
& \textbf{3-way Complement Merge}  & 26.33 & 99.43 \\
\midrule
\multirow{7}{*}{\textbf{MedQA}}
& Task-only               & 60.00  & - \\
& 2-step Finetuning       & 49.56  & 99.60 \\
& Same Merge              & 21.56  & 98.70 \\
& TrojanPlugin FUSION Merge & 57.75  & 98.47 \\
& \textbf{FF-only Merge}  & 53.32  & 98.72 \\
& \textbf{2-way Complement Merge} & 58.42  & 91.63 \\
& \textbf{3-way Complement Merge} & 57.45  & 99.36 \\

\bottomrule
\end{tabular}
}
\vspace{0.5em}
\end{table*}

\end{document}